\documentclass{egpubl}
\usepackage{pg2018}

\SpecialIssuePaper         
 \electronicVersion 

\ifpdf \usepackage[pdftex]{graphicx} \pdfcompresslevel=9
\else \usepackage[dvips]{graphicx} \fi

\PrintedOrElectronic

\usepackage{t1enc,dfadobe}

\usepackage{egweblnk}
\usepackage{cite}

\usepackage{amsmath}
\usepackage{amssymb}
\usepackage{subfigure}
\graphicspath{{figures/}}

\title[StretchDenoise: Parametric Curve Reconstruction]%
      {StretchDenoise: Parametric Curve Reconstruction with Guarantees by Separating Connectivity from Residual Uncertainty of Samples}

\author[S. Ohrhallinger \& M. Wimmer]
{\parbox{\textwidth}{\centering S. Ohrhallinger$^{1}$
        and M. Wimmer$^{1}$ 
        }
        \\
{\parbox{\textwidth}{\centering $^1$TU Wien, Austria
       }
}
}

\begin{document}

\teaser{
 \centering
\subfigure[Original smooth curve]{\includegraphics[width=1.6in]{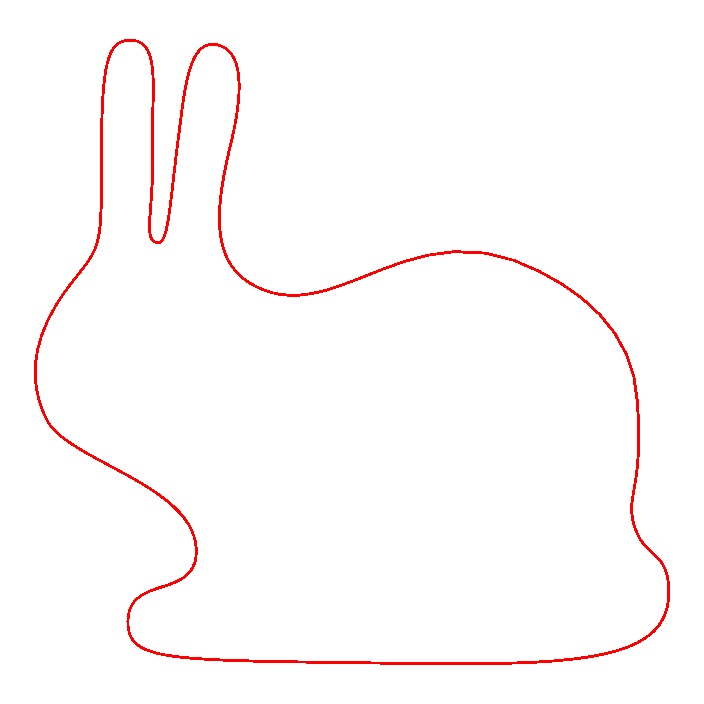}}\hfill
\subfigure[Samples with noise extent]{\includegraphics[width=1.6in]{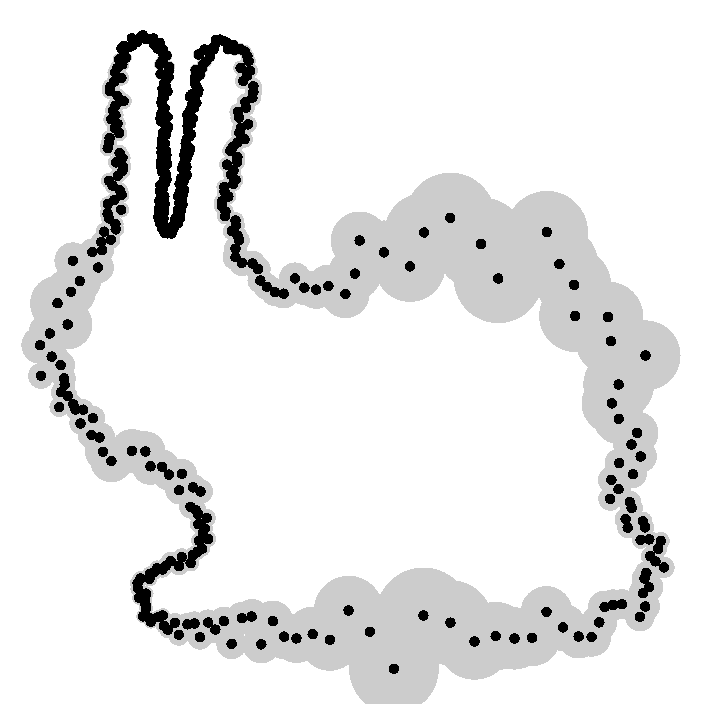}}\hfill
\subfigure[Pass \#1: Connected manifold]{\includegraphics[width=1.6in]{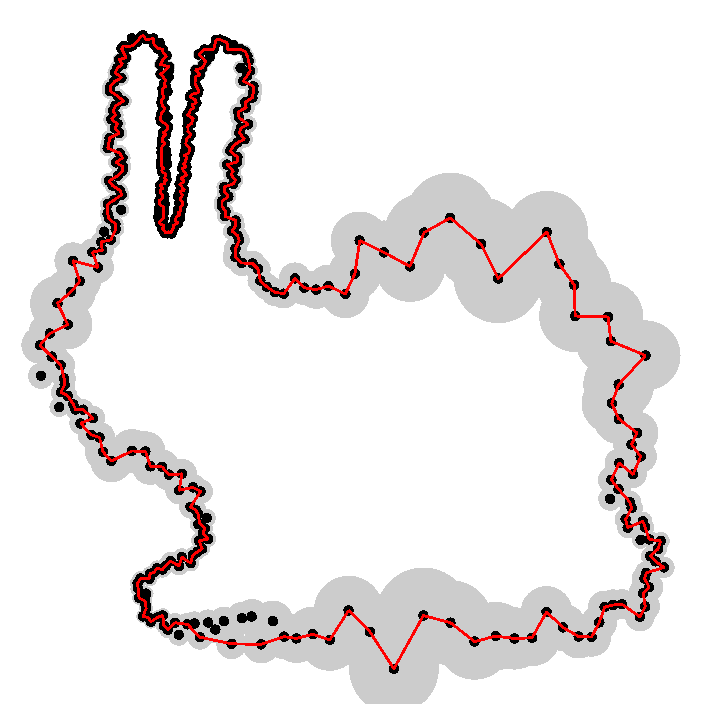}}\hfill
\subfigure[Pass \#2: Denoised curve]{\includegraphics[width=1.6in]{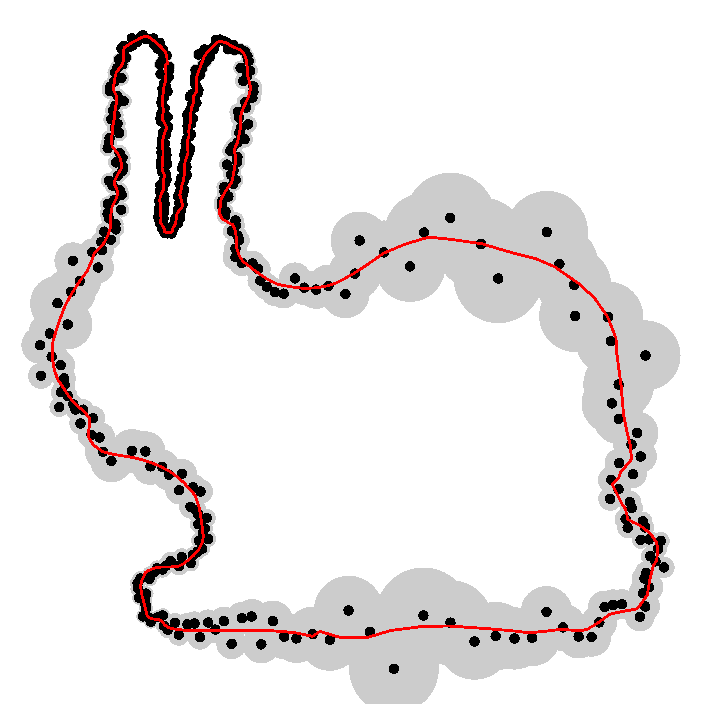}}\hfill
\caption{Our parameter-free method reconstructs features while effectively removing noise by a two-pass approach.}
\label{fig:teaser}
}

\maketitle
\begin{abstract}
We reconstruct a closed denoised curve from an unstructured and highly noisy 2D point cloud.
Our proposed method uses a two-pass approach: Previously recovered manifold connectivity is used for ordering noisy samples along this manifold and express these as residuals in order to enable parametric denoising.
This separates recovering low-frequency features from denoising high frequencies, which avoids over-smoothing.
The noise probability density functions (PDFs) at samples are either taken from sensor noise models or from estimates of the connectivity recovered in the first pass.
The output curve balances the signed distances (inside/outside) to the samples.
Additionally, the angles between edges of the polygon representing the connectivity become minimized in the least-square sense.
The movement of the polygon's vertices is restricted to their noise extent, i.e., a cut-off distance corresponding to a maximum variance of the PDFs. 
We approximate the resulting optimization model, which consists of higher-order functions, by a linear model with good correspondence.
Our algorithm is parameter-free and operates fast on the local neighborhoods determined by the connectivity.
We augment a least-squares solver constrained by a linear system to also handle bounds.
This enables us to guarantee stochastic error bounds for sampled curves corrupted by noise, e.g., silhouettes from sensed data, and we improve on the reconstruction error from ground truth.

Open source to reproduce figures and tables in this paper is available at: {\em https://github.com/stefango74/stretchdenoise}

\ccsdesc[500]{Computing methodologies~Shape modeling}
\ccsdesc[500]{Computing methodologies~Point-based models}

\printccsdesc   
\end{abstract}

\section{Introduction}

Reconstructing closed curves from noisy samples is considered an important problem in computational geometry by itself.
Furthermore it has applications in image analysis, computer vision and reverse engineering.
An example use case is the extraction of silhouettes from sensed depth images, which consist of noisy points, to segment the color data once reconstruction and denoising have generated clear contours.
Existing curve reconstruction and denoising methods often rely on Gaussian smoothing, which creates nice visual output but may oversmooth features.
Also the actual noise extent is not considered, even if sensor device properties are known, in order to (stochastically) guarantee the error of acquisition.

State-of-the-art curve reconstruction algorithms operating on noisy samples can estimate an extent of local noise for applying, e.g., Gaussian smoothing.
However, recovering the connectivity requires estimating the extent of noise, and the high frequencies of the signal, the noise, can in turn only be estimated well if the baseline of the signal, the connectivity, is known.
This mutual dependency is why such algorithms often output curves which are not manifold, or over-smooth features.
We therefore propose a two-pass approach:

First, to break up the mutual dependence of connectivity and noise, we apply {\scshape FitConnect}~\cite{ohrhallinger2018fitconnect}, an algorithm which manages to reconstruct the connectivity by testing for consistent manifold fittings of circular arcs as curve segments on increasing scales.
For a closed curve, it outputs a polygon with samples as vertices that are sparsely chosen in proportion of the size of noise clusters and therefore recover features.
These vertices are augmented with normals, and the neighborhood of samples contributing to its local curve fit.
This allows us to order and associate the noisy samples along the reconstructed connectivity, in a single-parametric space, with their Hausdorff distances as residuals separated from the underlying low-frequency manifold connectivity.

Secondly, we move the vertices of the reconstructed polygon to find the {\em most probable curve} fitting the noisy samples.
We maximally straighten the curve while keeping it within the error bounds, specified based on sensor noise models, for example. If a cut-off PDF is used, a probability of being within the ground truth can be guaranteed. 
At the same time we keep the samples' Hausdorff distances balanced between the in- and outside of the curve to avoid area shrinking.

Our contributions are:

\begin{itemize}
\item A two-pass reconstruction approach that uses prior connectivity to {\em enable a simpler and more efficient denoising model} while {\em conserving features emerging over the noise extent} (see Figure~\ref{fig:teaser}).
\item A {\em parameter-free} denoising method with {\em stochastic guarantees}.
\item {\em A constrained least-squares solver that can handle bounds}.
\end{itemize}

\section{Related Work}

First, we take a look at the state of the art for reconstructing curves from noisy point sets and denoising noisy curves, and their applications.

\textbf{Applications for reconstructing curves from noisy samples}
Birkas et al.~\cite{birkas2016mobile} take sensed RGBD images and cluster points to extract silhouettes.
With the reconstructed silhouette curve, the corresponding object can be segmented and visualized in the RGB part of the image.
However, these point sets are polluted by high noise from the mobile sensor and for exact segmentation, a denoised curve is required.
The probability density functions of that noise has been analyzed for different sensor devices~\cite{koeppel-2017-baa,grossmann-2017-baa}.

\textbf{Curve reconstruction from noisy samples}
The method of Lee~\cite{lee2000curve} uses a neighborhood graph, the {\em Euclidean Minimum Spanning Tree}, to connect noisy samples.
It then smoothens this thick graph using a variant of {\em Moving Least Squares}~\cite{levin1998approximation} and applies a spline fit.
Their method is limited to single open curves and does not handle varying sample density or noises well.
{\em Screened Poisson}~\cite{kazhdan2013screened} relies on given normals for noisy point cloud reconstruction~\cite{alliez2007voronoi}, however normals from sensor input are often noisy as well.
{\em Robust HPR}~\cite{mehra2010visibility} extracts connectivity locally from a transformation of the convex hull and combines it in a weighted global graph.
However, it often exhibits gaps in the reconstruction and does not produce a denoised curve.
~\cite{degoes2011optimal} solve a related problem and can also reconstruct intersecting curves by greedily simplifying a Delaunay triangulation of the point set but fail to connect curves with non-uniform sampling or noise.
A related method~\cite{wang2014robust} also fails for non-uniform sampling.
One method~\cite{rupniewski2014curve} moves and eliminates balls centered on samples to obtain a sparse piece-wise linear fit but shows results only on simple cases with very dense sampling.
The recent method {\scshape FitConnect}~\cite{ohrhallinger2018fitconnect} fits circular neighborhoods, as has been shown to work well~\cite{guennebaud2007algebraic}, and determine the inside/outside of the curve locally.
{\scshape FitConnect} increases the neighborhood size for the fits until they become consistent with each other, eliminating samples in the process that do not contribute to the connectivity.
They guarantee manifold construction for arbitrarily high noise, provided that the features emerge over the noise extent, and provide an estimate of the local noise at samples.
This algorithm is a direct extension of a reconstruction algorithm handling noise-free samples under very relaxed conditions~\cite{ohrhallinger2016hnn}, which also gives a detailed overview of prior such work as important groundwork in this field.
As a post-processing, {\scshape FitConnect} blends the locally determined neighborhood fits along normals, but this can result in jagged edges where fits vary much in size, or where it interpolates samples where the noise extent is too low to be detected.
We will use the connectivity reconstruction algorithm of {\scshape FitConnect}~\cite{ohrhallinger2018fitconnect} and apply a denoising algorithm based on the properties recovered together with that connectivity. As a result, we are able to both denoise interpolated points with a specified noise extent and avoiding jagged edges.

\textbf{Guarantees for curve reconstruction}
Dey and Goswami~\cite{dey2006provable} describe a noise model that expresses the noise at samples in terms of their local feature size.
Without quantifying that fraction, they prove that reconstruction is, in principle, possible.
Cheng et al.~\cite{cheng2005curve} prove that this reconstruction is possible with a probability in terms of a function of noise at samples and the local feature size, however, their proposed algorithm is of unpractical $O(N^3)$ time complexity for a number of $N$ points.
Also, it requires locally uniform distribution and uniform perturbation in the normal directions.
We are using {\scshape FitConnect}~\cite{ohrhallinger2018fitconnect} as base for recovering the connectivity, which was shown to reconstruct features that locally emerge over the extent of noise at the samples.

\textbf{Curve denoising}
There are various approaches to denoising an existing curve. One method fits a boundary to regions with noisy points and then applies region thinning~\cite{song2010boundary}.
But since this relies on area instead of considering the density and contribution of samples, it will not produce correct results for varying sampling densities.
Another method~\cite{feiszli2011curve} applies multi-scale analysis using a Gaussian kernel but preserves sharp points with a shock detector by defining a model of the output curve as a collection of smooth arcs and corners.
A further method~\cite{lu2015rapid} uses Gaussian smoothing for noise estimated by local analysis, with fixed $n=30$ neighbors.
This will, like similar methods, over-smooth features in regions of the point set that are not highly noisy.
Additionally, known noise extents, e.g., from sensed data, are not considered by these algorithms.
In our method, we can specify these noise extents at samples to give a stochastic guarantee of reconstruction distance to the original curve.

\textbf{Constrained optimization techniques}

For our denoising, we need to solve a constrained least-squares minimization problem~\cite{coons1978constrained,lawson1995solving} that is not only constrained by linear equalities, which can be solved using Lagrangian multipliers~\cite{selesnick2013constrainedls}, but also by bounds, which is also closely related to linear programming~\cite{kantorovich1940new}.
Since none of these methods is able to directly solve our model, we design our own variant.

\section{Problem Definition}

As input we take a set of noisy points $S$ sampling a closed smooth curve $C$.
We obtain the connectivity by running the algorithm {\scshape FitConnect}~\cite{ohrhallinger2018fitconnect}, which fits a linear piece-wise curve to the samples, i.e., a polygon $P$ with vertices $V \subseteq S$.
To do so, {\scshape FitConnect} iteratively fits increasing $k$-neighborhoods of noisy samples with circular arcs until adjacent fits become mutually consistent.
In that process it eliminates samples in noisy clusters which are redundant w.r.t. connectivity.
For the remaining points it blends the arcs along their determined normals as a simple post-processing step to approximate the original curve. In this paper, we omit this step in order to apply our own denoising method, which assumes the following input:
Each vertex $v_i \in V$ has a neighborhood $N_i$, which is a list of samples in $S$ ordered by their projection onto its fit, as well as a normal $n_i$ and a maximum noise extent $r_i$ detected by {\scshape FitConnect} ($r_i$ is zero if the sample can be interpolated without requiring fitting to local noise).
In case a noise cut-off radius $r_i$ is available from another source, e.g., if a sensor noise model is known, we will take these values as input instead.
With $d(x,P)$ being the Hausdorff distance between a point $x$ and polygon $P$, we define its signed variant as:
\begin{equation}
\hat{d}(x)=\begin{cases}d(x,P), & \text{if $x$ on or outside $P$}.\\-d(x,P), & \text{if x inside P}.\end{cases}
\end{equation}


Noise from sensed data is often modeled as a Gaussian probability distribution function (PDF).
In our use case -- silhouettes extracted from sensed data and projected onto the view plane as point sets -- we only consider lateral noise and define a simplified isotropic radial PDF, since this corresponds closely to the x- and y-axis distribution of sensed data~\cite{koeppel-2017-baa,grossmann-2017-baa}:
\begin{equation}
f_X(x) =\frac{1}{\sigma\sqrt{2\pi}}\exp\left\{-\frac{(x-\mu)^2}{2\sigma^2}\right\}, \sigma>0
\end{equation}
This guarantees the sample to lie within a cut-off radius $r$ with probability $\Pi$, which depends on a user-defined maximum allowed $\sigma$. 

\begin{figure}
\centering
\includegraphics[width=3in]{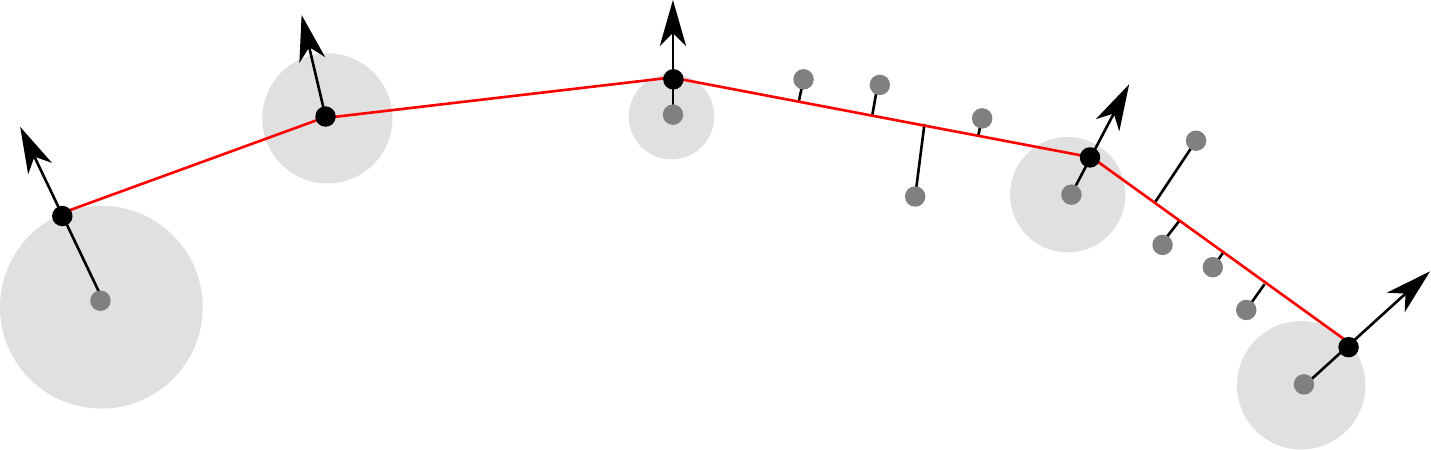}
  \caption{Our denoising goals: Grey dots are noisy samples $s_i$, black dots are final vertices $v'_i$ of the red curve polygon $P'$ which can move along their normals. 1) Regularizing the curve by least-square minimizing of angles $\alpha_i$ between adjacent edges. 2) Balance the curve such that the total signed distance of samples to their nearest edge $\hat{d}(s_i,e)$ equalizes to zero. 3) Keep the curve vertices inside the discs of noise extent $r_i$.}
\label{fig:goals}
\end{figure}

To achieve a curve that optimally both denoises and fits the noisy samples, we pursue the following three goals (see Figure~\ref{fig:goals}):

\begin{enumerate}
\item \textbf{Eliminate high frequencies (noise)} by regularizing the curve in the sense of straightening it where no features protrude over the noise extent. We achieve this maximal denoising of the curve by minimizing the angles of the polygon in the least-squares sense:
\begin{equation}
\label{eq:ls-angles}
\arg\min_V \sum \|\alpha_i\|_2^2, \alpha_i=\angle \overrightarrow{v_{i-1},v_i}, \overrightarrow{v_i,v_{i+1}}
\end{equation}
\item \textbf{Balancing the curve} with respect to the number of samples that lie inside and outside. This is achieved by setting the desired mean signed distance to $P$ to zero:
\begin{equation}
\label{eq:sd-balance}
\sum\limits_i^{|S|} \hat{d}(s_i)=0.
\end{equation}
Using the signed distance prohibits area shrinking.
\item \textbf{Bounding the curve} within the discs $D_i(v_i, r_i$) of the maximum permitted distance from samples, in order to preserve the features recovered by {\scshape FitConnect}:
\begin{equation}
\{\forall s_i \in S: d(s_i,P)\leq r_i\}
\end{equation}
This results in the stochastic guarantee of the samples having been produced by the curve with probability $\Pi$.
\end{enumerate}

Note that we do not consider outlier points, for example introduced by sensing errors.
Those are not connected to $P$ by {\scshape FitConnect} since they lie too far from the curve to be mutually consistent with inlier points. Thus, we assume $V$ to be free of outliers.

\section{Denoising Algorithm}

The above-mentioned constrained optimization model poses some challenges: It allows too much freedom, and is formulated globally, both of which make it difficult to solve it effectively and in reasonable run time.
Moving the polygon vertices $V$ freely in $\mathbb{R}^2$ would result in higher-order functions in the minimization problem as well as in the constraints and bounds, making it slow to solve and becoming trapped inside local minima.
Since the curve polygon is locally mostly tangential to the normals anyway, free movement is too lenient and we restrict the problem by allowing vertices $v_i$ to move only along their normals $n_i$.
This allows us to model all functions as linear ones, enabling fast solving for the minimum, and we do not expect a significant deviation from the minimum of the exact model specified above. 

\subsection{Adapted Model}

\begin{figure}
\centering
\includegraphics[width=3in]{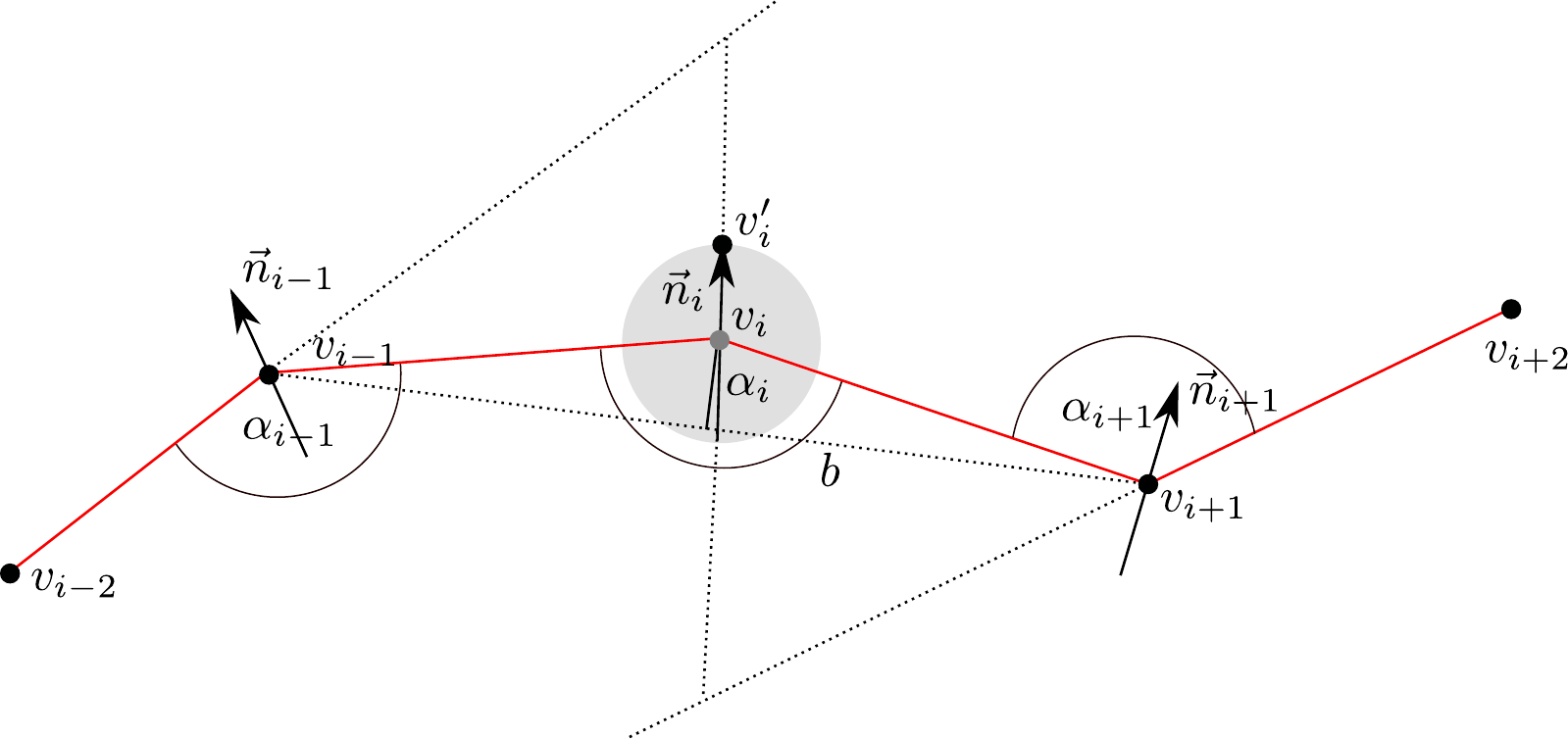}
  \caption{The angle $\alpha_i$ is approximated by the distance of $v_i$ to edge $\mathbf{b}$ between its adjacent vertices, weighted by its inverse length. Moving $v_i$ along $n_i$ also changes the adjacent angles $\alpha_{i-1},\alpha_{i+1}$ with a factor of the dot product of their associated normals.}
\label{fig:model-angles}
\end{figure}

We adapt and detail the above-mentioned model in the following ways to obtain linear functions:

Let $\mathbf{v}'_i=\mathbf{v}_i+x_i\mathbf{n}_i$, with $x_i \in \mathbf{x}$ as a vector of displacement scalar values and $\mathbf{n}$ as the normalized normals at $\mathbf{v}$.


\begin{enumerate}
\item \textbf{Angles:} We approximate the non-linear computation of an angle between incident edges of a vertex $\mathbf{v}'_i$ by its linear distance to the baseline $\mathbf{b}$ of its neighbor vertices, weighted by its reciprocal length to get relative values proportional to angles:
\begin{equation}
y(i)=\frac{d(\mathbf{v}_i, \mathbf{b})}{\|\mathbf{b}\|}, \mathbf{b}=(\mathbf{v}_{i-1},\mathbf{v}_{i+1}),\approx\alpha_i=\angle (\mathbf{v}_{i-1},\mathbf{v}_i),(\mathbf{v}_i,\mathbf{v}_{i+1})
\end{equation}
Both angle and the weighted distance correspond at their zero values.
Since these values are summed up as squares before minimizing, we expect the non-linear mapping to have little impact.
When we move a $\mathbf{v}_i$ to $\mathbf{v}'_i=\mathbf{v}_i+x_i\mathbf{n}_i$, this affects not only $\alpha_i$ but also adjacent $\alpha_{i-1}$ and $\alpha_{i+1}$, multiplied by the dot product of their normals $\mathbf{n}_{i-1}$,$\mathbf{n}_{i+1}$ with $\mathbf{n}_i$ (see Figure~\ref{fig:model-angles}),
and therefore:
\begin{equation}
H(i-1,i)=\mathbf{n}_{i-1}^T\mathbf{n}_i\frac{d(\mathbf{v}'_i, (\mathbf{v}_{i-2},\mathbf{v}_{i-1}))}{\|\mathbf{b}\|}
\end{equation}
\begin{equation}
H(i,i)=\mathbf{n}_{i-1}^T\mathbf{n}_i\frac{d(\mathbf{v}'_i, \mathbf{b})}{\|\mathbf{b}\|}
\end{equation}
\begin{equation}
H(i+1,i)=\mathbf{n}_i^T\mathbf{n}_{i+1}\frac{d(\mathbf{v}'_i, (\mathbf{v}_{i+1},\mathbf{v}_{i+2}))}{\|\mathbf{b}\|}
\end{equation}
We can then substitute into Equation~\ref{eq:ls-angles} to approximately express the linear squares minimization of angles in terms of $\mathbf{x}$:
\begin{equation}
\label{eq:ls-dist}
\arg\min_\mathbf{x} \|\mathbf{Hx}-\mathbf{y}\|_2^2
\end{equation}
as a sparse diagonal matrix with 3 non-zero colums per row.

\item \textbf{Balance:} When we move a vertex $v_i$, this displaces its two adjacent edges $e_{i,prev}(v_{i-1},v_i)$ and $e_{i,next}(v_i,v_{i+1})$.
In turn, this affects the Hausdorff distance of the samples $S_e$ closest to an edge $e$.
We consider the initial distance of samples as orthogonal to the edge:
\begin{equation}
\label{eq:sample-distance}
b_i(\mathbf{e})=\sum\limits_{s_j}^{S_e} (\mathbf{s}_j-\mathbf{v}_i)^T \mathbf{n_e}, \mathbf{n_e}=\bot\mathbf{e}
\end{equation}
and clamped unit values of samples' positions along the edge since they will move more in terms of $x_i$ the closer they are to $v_i$, with a factor of $[0,1]$:
\begin{equation}
\label{eq:sample-position}
c_i(\mathbf{e})=\sum\limits_{s_j}^{S_e} \frac{(\mathbf{s}_j-\mathbf{v}_i)^T \mathbf{e}}{\|\mathbf{e}\|^2}|_{[0,1]}
\end{equation}
so that we can express the displacement of samples in terms of $x_i$ along $n_i$ approximatively by substituting Equations~\ref{eq:sample-distance} and~\ref{eq:sample-position} into Equation~\ref{eq:sd-balance}. This computes the distances of the samples $x_i c_i$ from the moving edge minus their initial displacement $b_i$:
\begin{equation}
\label{eq:linear}
\sum\limits_i^{|S(v_i)|} x_i[c_i(\mathbf{e}(s_i))]-b_i(\mathbf{e}(s_i))=0
\end{equation}
Note that while our initial (constant) displacement corresponds to the Hausdorff distance as being orthogonal to the edge, we use distance along the vertex normal to approximate this quadratic term by a linear one. 
Since the linear term (non-orthogonal distance of point to line) is an upper bound of the quadratic term (Hausdorff distance), $x_i$ values will not diverge.

\item \textbf{Bounds:} We set lower and upper bounds:
\begin{equation}
\label{eq:bounds}
\{\forall i \in |S|: -r_i \leq x_i \leq r_i\}
\end{equation}
Note that this would also permit using anisotropic PDFs.
\end{enumerate}

Our adapted model now contains:
\begin{itemize}
\item A least-squares minimization (Equation~\ref{eq:ls-dist})
\item A linear system (Equation~\ref{eq:linear}) with a single row and
\item Lower+upper bounds (Equation~\ref{eq:bounds}).
\end{itemize}

Concisely we formulate this as:
\begin{equation}
\begin{split}
\text{minimize } \mathbf{Hx-y} \\
\text{subject to } \mathbf{Cx-b=0}\\
\text{and } \mathbf{-r \leq x \leq r}
\end{split}
\end{equation}

\subsection{Our Augmented Solver}

We are not aware of a technique to solve this bound-constrained optimization problem directly.
{\em Linear programming}~\cite{kantorovich1940new} is a popular method to minimize an objective function with bounds, and also supports (in)equality constraints.
However, the objective function must consist of a single row, but we need to minimize independently per angle, and at once, so we have a matrix with multiple rows.
We therefore apply {\em constrained least squares}~\cite{coons1978constrained,lawson1995solving} to solve for the first two equations, and then augment that technique with bounds. By defining Lagrangian multipliers for the linear equation $\mathbf{Cx-b=0}$ and setting its derivatives to zero we can transform:

\begin{equation}
\min_\mathbf{x} \|\mathbf{Hx}-\mathbf{y}\|_2^2 \text{s.t.} \mathbf{Cx-b=0}
\end{equation}

into (see~\cite{selesnick2013constrainedls}):

\begin{equation}
\mathbf{x}=(\mathbf{H}^T\mathbf{H})^{-1}(\mathbf{H}^T\mathbf{y}-\mathbf{C}^T(\mathbf{C}(\mathbf{H}^T\mathbf{H})^{-1}\mathbf{C}^T)^{-1}(\mathbf{C}(\mathbf{H}^T\mathbf{H})^{-1}\mathbf{H}^T\mathbf{y-b}))
\end{equation}

Solving this expression may result in values $x_i \in \mathbf{x}$ which violate the bounds ($\mathbf{-r,r}$).
To incorporate the bounds in this solver, we first clamp each out-of-bound value $x_i$ to its respective (lower or upper) bound.
Then, we treat that clamped $\hat{x}_i$ value as constant and eliminate its corresponding column from both $\mathbf{H}$ and $\mathbf{C}$ by substituting the eliminated values into $\mathbf{y}$ and $\mathbf{b}$ respectively.
We iterate until all $x_i$ are either inside their bounds or have become constant.

Note that our model considers the bounds from noise cut-off radii only at vertices, not at all samples.
We could implement bounds-checking also per-sample by checking all associated samples of incident edges per vertex, but omit it since noise extent at vertices is representative for the associated neighborhood. 

\subsection{Solving Locally}

If we apply this solving technique to the entire curve polygon at once, it would be quite slow, since the required matrix operations are of super-quadratic time complexity in the number of vertices, even if that number is just proportional to the count of features, not of samples.
Our experiments also showed that balancing the curve inside/outside globally can result in directional shifts, as displacements equalize out over the varying orientations, which is not desired.


For these reasons we apply our solver to more fine-grained subsets separately, such that they are large-scale enough to remove the noise but still so local as to avoid this shifting effect.
We determine these local subsets by starting at an initial vertex and adding adjacent vertices in both directions while a line intersects all discs of their noise extents.
Since a straight curve segment could fit all these vertices, we can eliminate its noise entirely without losing a feature, and since normals usually do not change orientations inside that subset, no shift will occur.
We continue along the polygon starting with the last affected vertex until all vertices have been visited and their displacements $x_i$ computed.

\textbf{Associating samples to edges of $P$}


As input from {\scshape FitConnect} we get for each vertex $v_i$ the neighborhood of samples $N_i$ (making up the consistent fit), ordered along their projection on the fitted circular arc for $N_i$. Samples can also be contained in multiple neighborhoods.
However, in order to find the Hausdorff distances of these samples to the polygon's edges, we need to locate for each sample the single closest edge.

Therefore, for each sample $s_j$, we analyze each of its containing neighborhoods $N_i(s_j)$.
For their $v_i$, we store the Hausdorff distance for each $s_j$ to both the preceding and successive edge of $v_i$.
We then associate $s_j$ to the closest edge among all containing $N_i(s_j)$.

\textbf{Testing whether adjacent vertices can be fit by a line}

In order to determine if a set of vertices $V_a \in V$ which are consecutive in $P$ can be fit by a line segment within their noise extents, we need to test whether there exists a line intersecting all the discs $D_i(v_i, r_i) \in D_a$, such that $D_a$ are the discs centered at the vertices $v_i \in V_a$, and with radius of their noise extent $r_i$.


For easier computation we use a non-affine transformation to map the discs into a unit space $\mathbb{U}[0,1][-1,1]$ by transforming the edge between the two boundary vertices to the unit line $l_u$ as the x-axis of $\mathbb{U}$ and scaling their radii to unit size 1 each.
Then we compute the top $t$ and bottom $b$ height of all inside discs $D_i$ w.r.t. $l_u$.
We determine the highest bottom value as $b_{max}=\max\limits_{\forall D_i \in D_a} b(D_i)$ and similarly, the lowest top value as $t_{min}=\min\limits_{\forall D_i \in D_a} t(D_i)$.
Now we select all discs $D_{above} \subseteq D_a$ with their bottom above the lowest top $t_{min}$ and all discs $D_{below} \subseteq D_a$ with their top below the highest bottom $b_{max}$.
 If both $D_{above}=\{\}$ and $D_{below}=\{\}$, there exists a line intersecting all of $D_a\setminus\{D_0,D_k\}$ which is parallel to $l_u$.
If also $t_{min}>-1.0$ and $b_{max}<1.0$ holds, this line intersects $D_0,D_k$ as well.
Else we have to test for lines not parallel to $l_u$:
For all discs $D_i \in D_{above}$, we construct the internal tangents $t_{ij}$ with all discs $D_j \in D_{below}$ and test if a tangent $t_{ij}$ exists which intersects all other discs $D_a \setminus \{D_i,D_j\}$.

\section{Results}

We have analyzed a large number and wide variety of point sets with our method.
This includes (1) data sets from related work in order to compare and show our improvements, (2) synthetic data sets to measure the reconstruction error with respect to ground truth in order to demonstrate the guarantees, and (3) real data, i.e., segmented silhouettes from noisy sensed data.
Open source code that replicates all result figures and tables of this paper is available online. 

\subsection{Improvements over prior work}

We compare our proposed method with three others that are able to reconstruct curves from actual point sets polluted by noise: the recent {\scshape FitConnect}~\cite{ohrhallinger2018fitconnect}, of which our method uses the connectivity reconstruction part; {\scshape Robust HPR}~\cite{mehra2010visibility}; and a method limited to open curves from Lee~\cite{lee2000curve}.


\textbf{Reconstruction error}

\begin{figure*}
\centering
\subfigure[$\delta$=0.1r]{\includegraphics[width=1in]{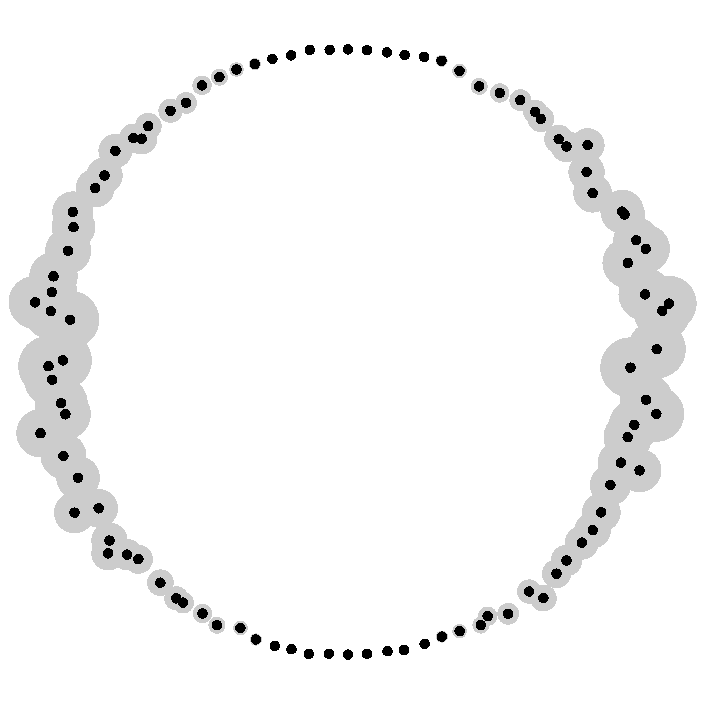}}\hfill
\subfigure[$\delta$=0.25r]{\includegraphics[width=1.1in]{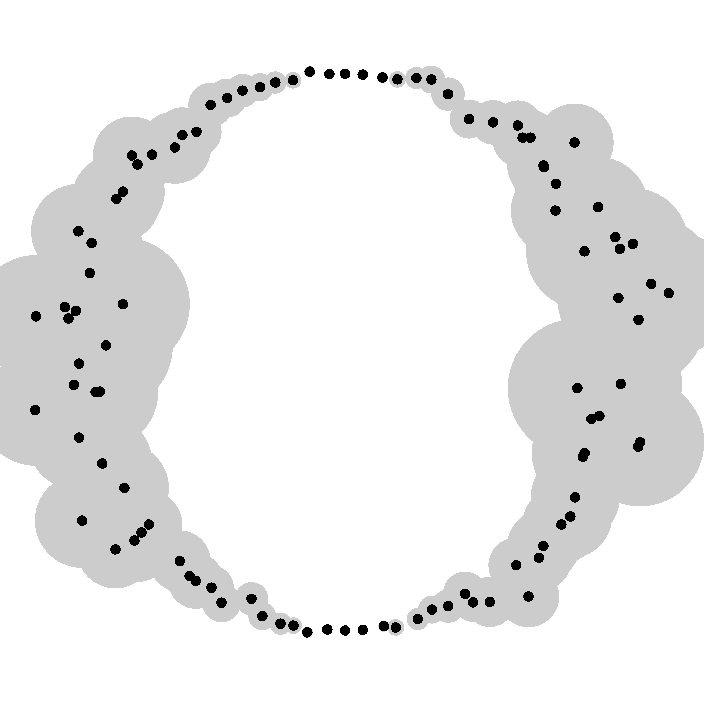}}\hfill
\subfigure[$\delta$=0.5r]{\includegraphics[width=1.25in]{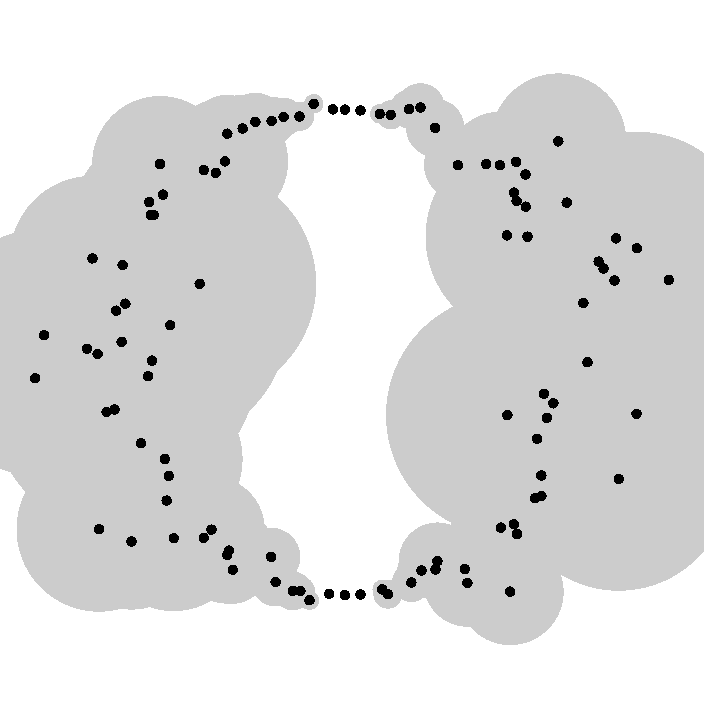}}\hfill
\subfigure[$\delta$=0.75r]{\includegraphics[width=1.4in]{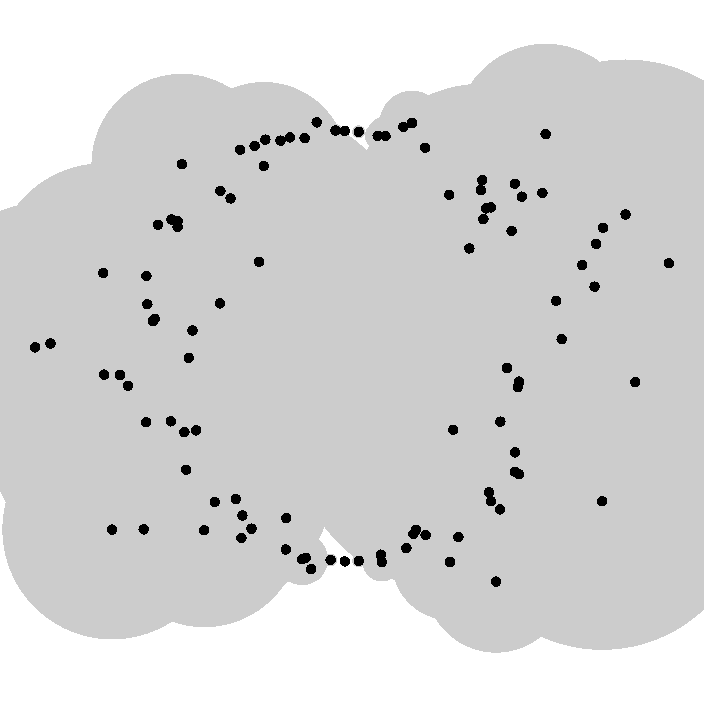}}\hfill
\subfigure[$\delta$=r]{\includegraphics[width=1.6in]{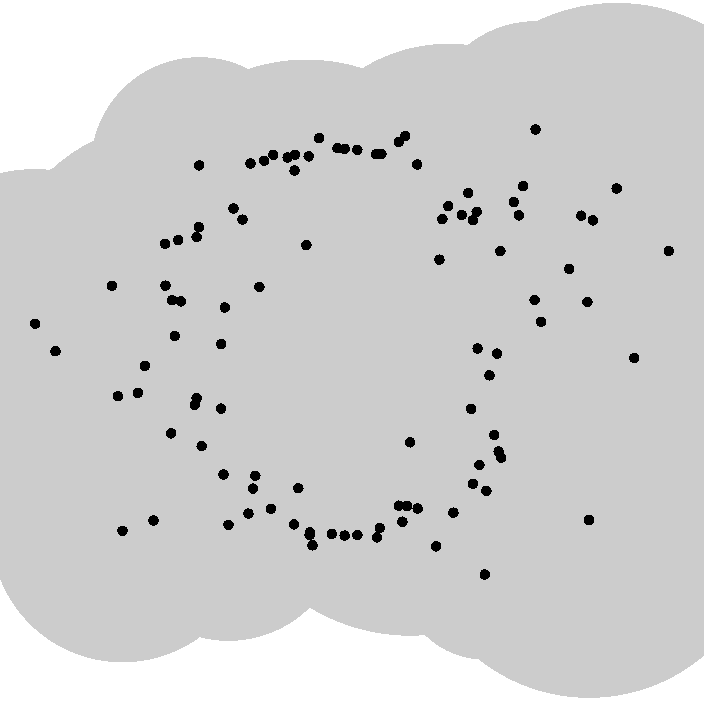}}
\subfigure[$\delta$=0.1r]{\includegraphics[width=1in]{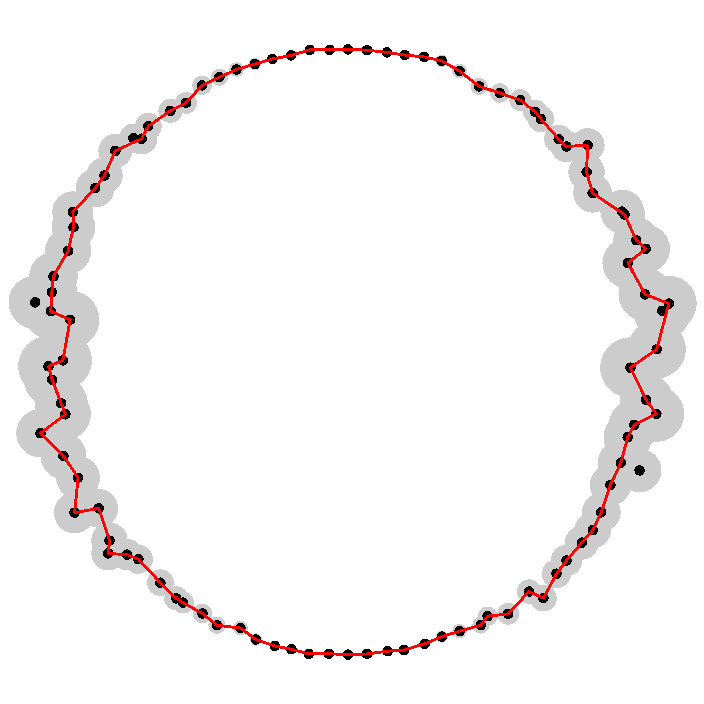}}\hfill
\subfigure[$\delta$=0.25r]{\includegraphics[width=1.1in]{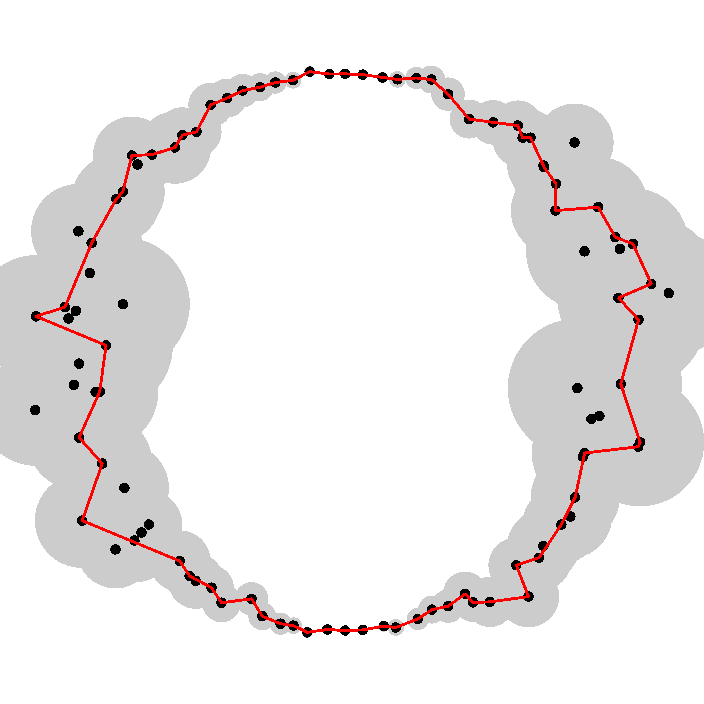}}\hfill
\subfigure[$\delta$=0.5r]{\includegraphics[width=1.25in]{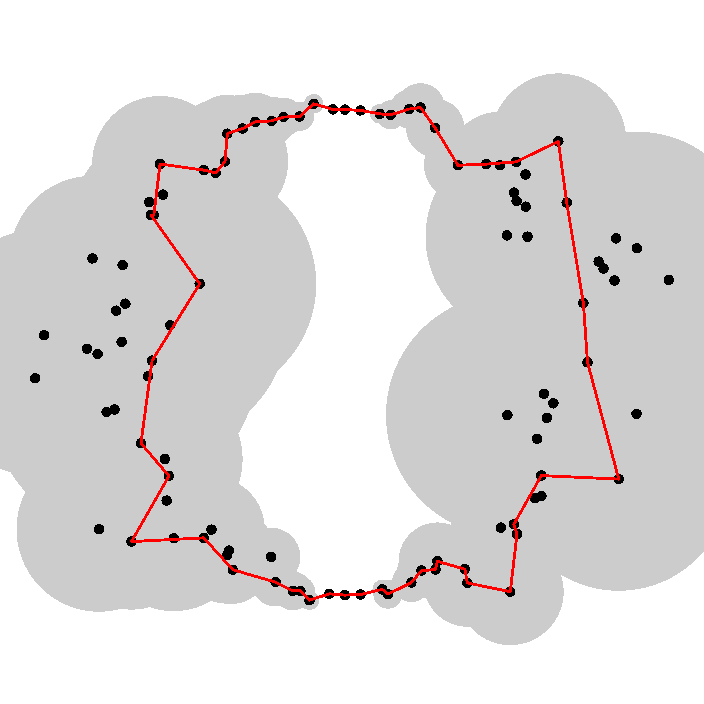}}\hfill
\subfigure[$\delta$=0.75r]{\includegraphics[width=1.4in]{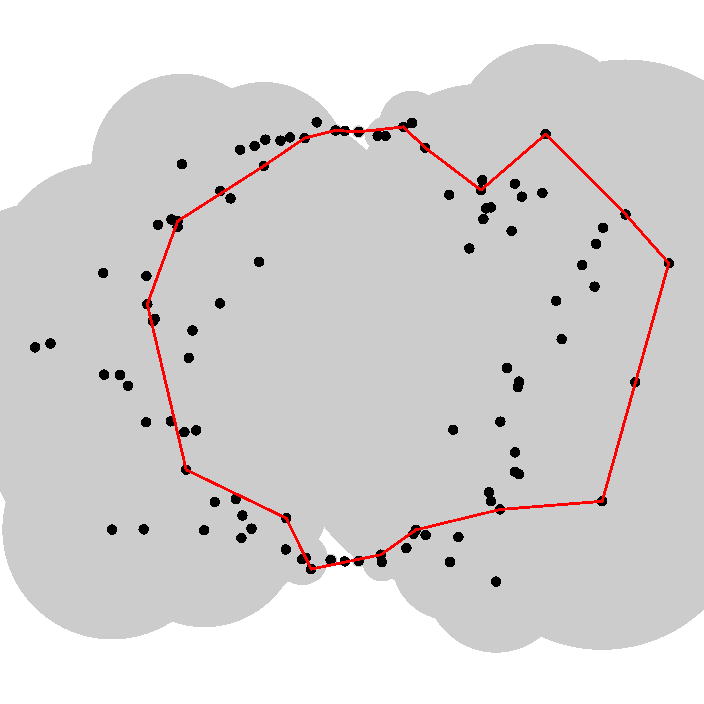}}\hfill
\subfigure[$\delta$=r]{\includegraphics[width=1.6in]{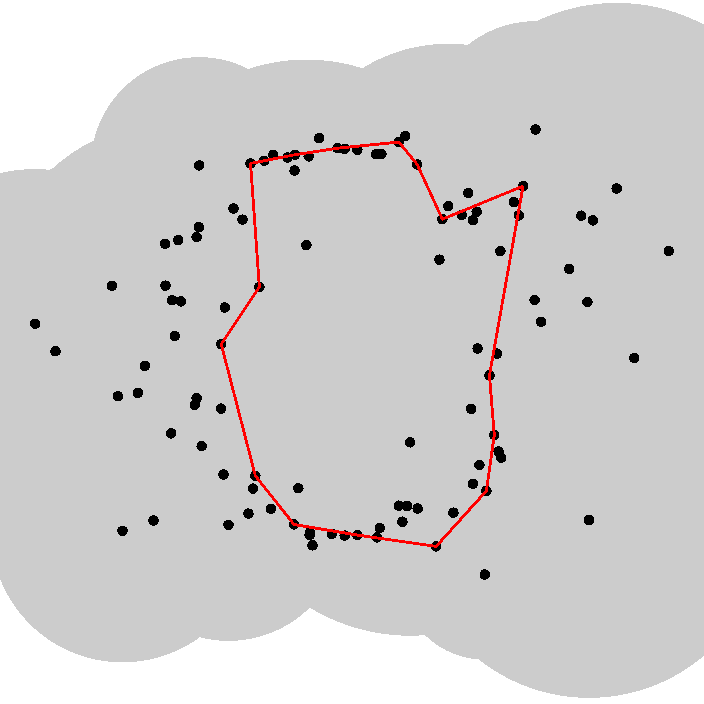}}
\subfigure[$\delta$=0.1r]{\includegraphics[width=1in]{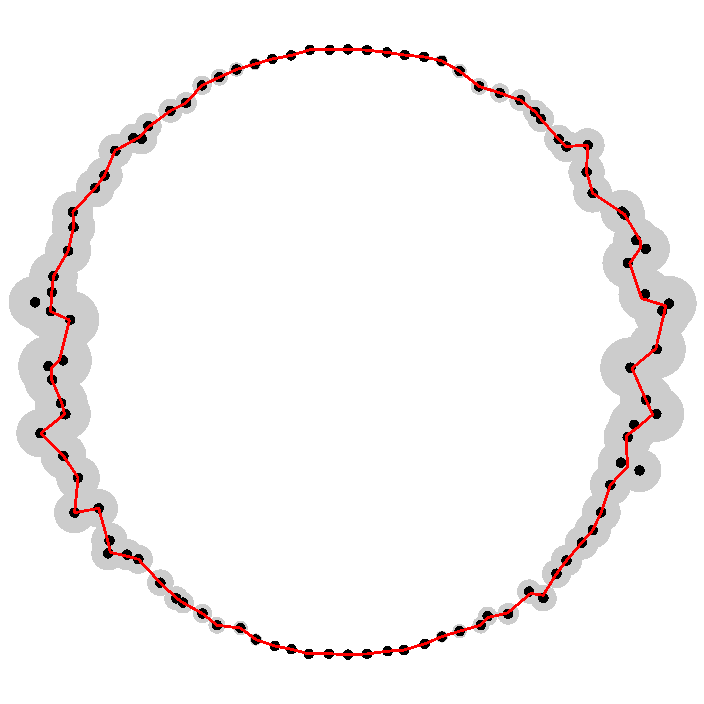}}\hfill
\subfigure[$\delta$=0.25r]{\includegraphics[width=1.1in]{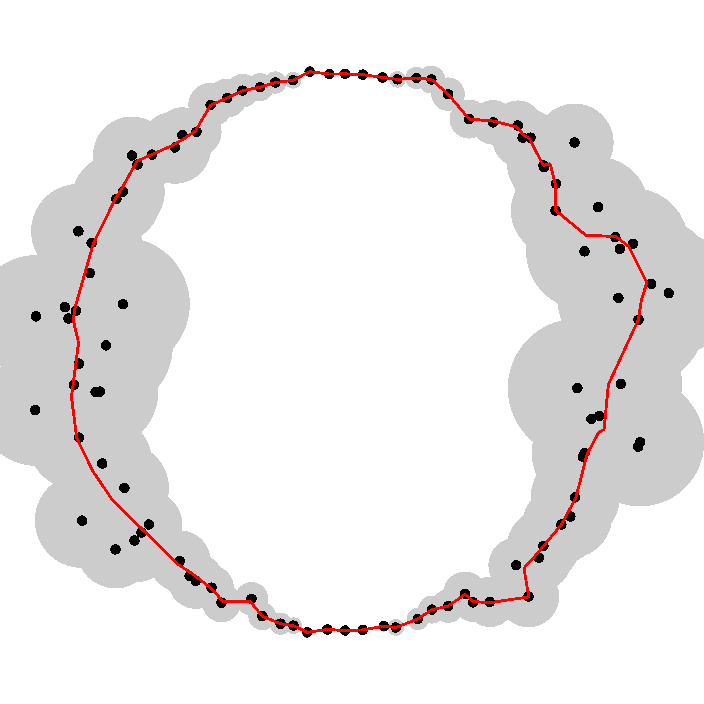}}\hfill
\subfigure[$\delta$=0.5r]{\includegraphics[width=1.25in]{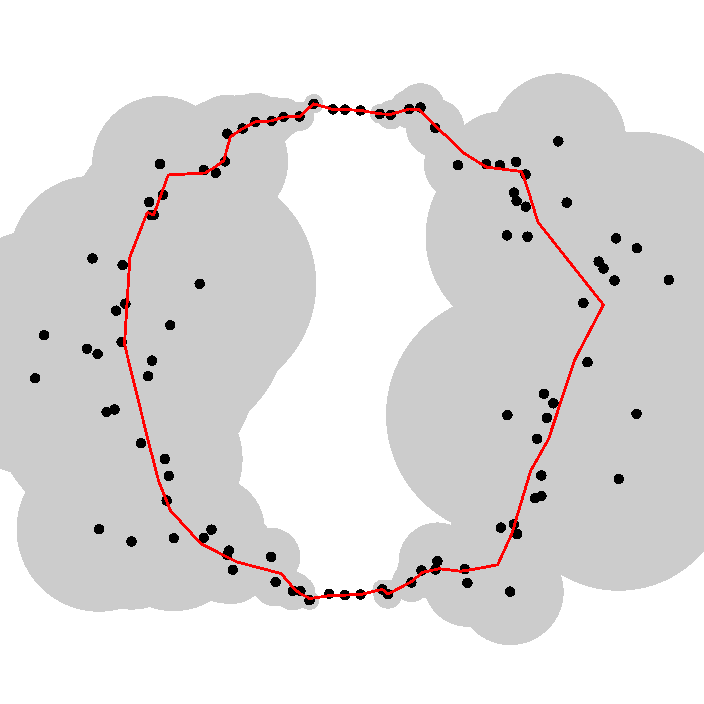}}\hfill
\subfigure[$\delta$=0.75r]{\includegraphics[width=1.4in]{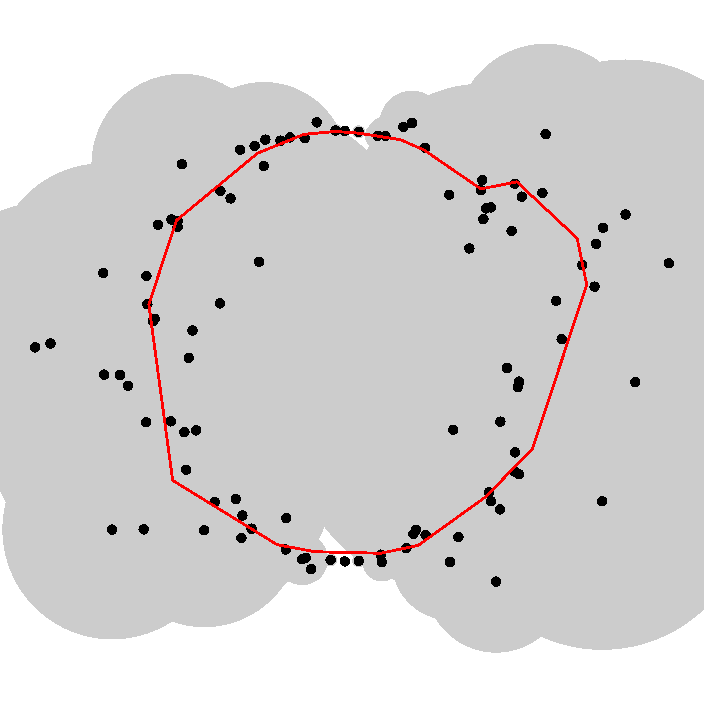}}\hfill
\subfigure[$\delta$=r]{\includegraphics[width=1.6in]{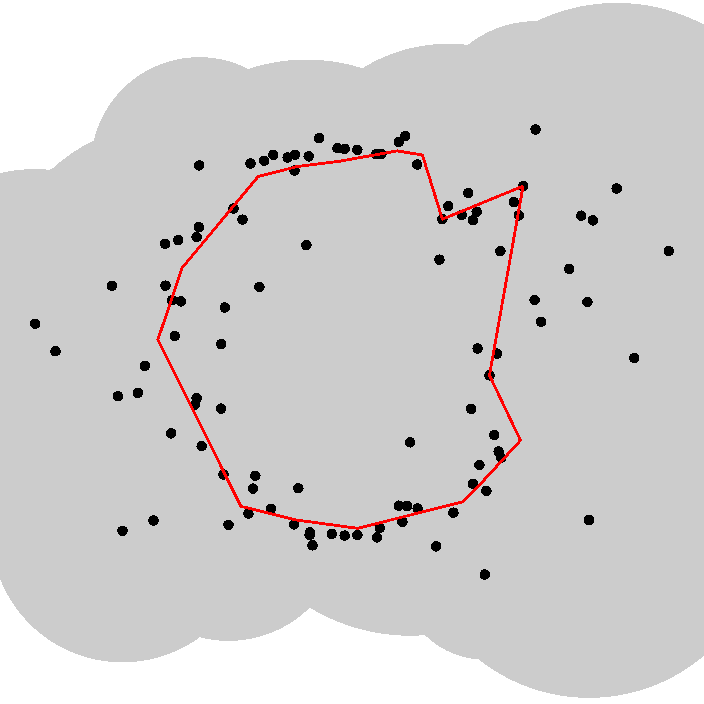}}
\subfigure[$\delta$=0.1r]{\includegraphics[width=1in]{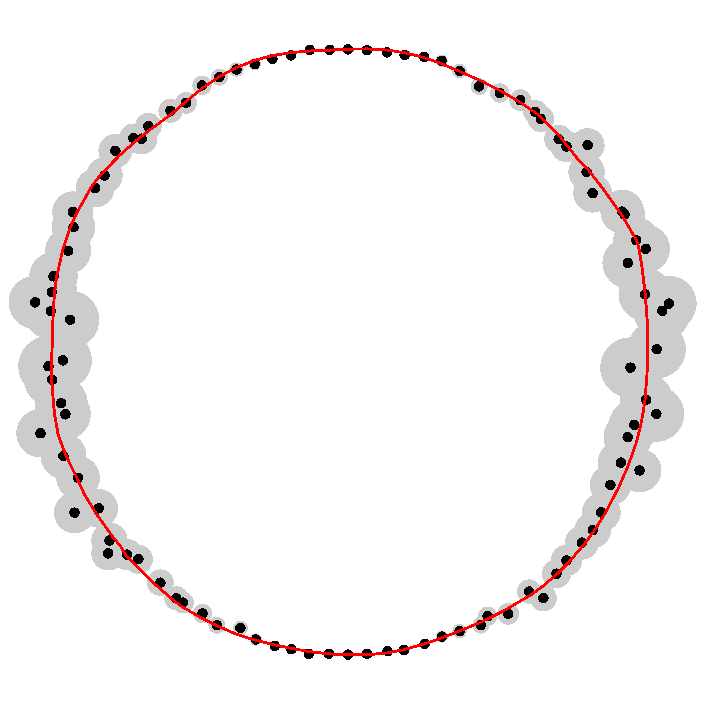}}\hfill
\subfigure[$\delta$=0.25r]{\includegraphics[width=1.1in]{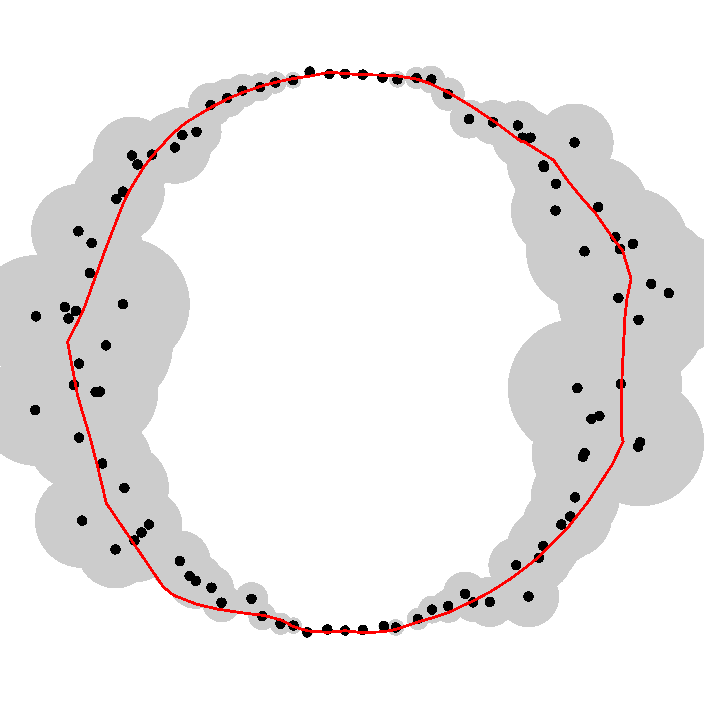}}\hfill
\subfigure[$\delta$=0.5r]{\includegraphics[width=1.25in]{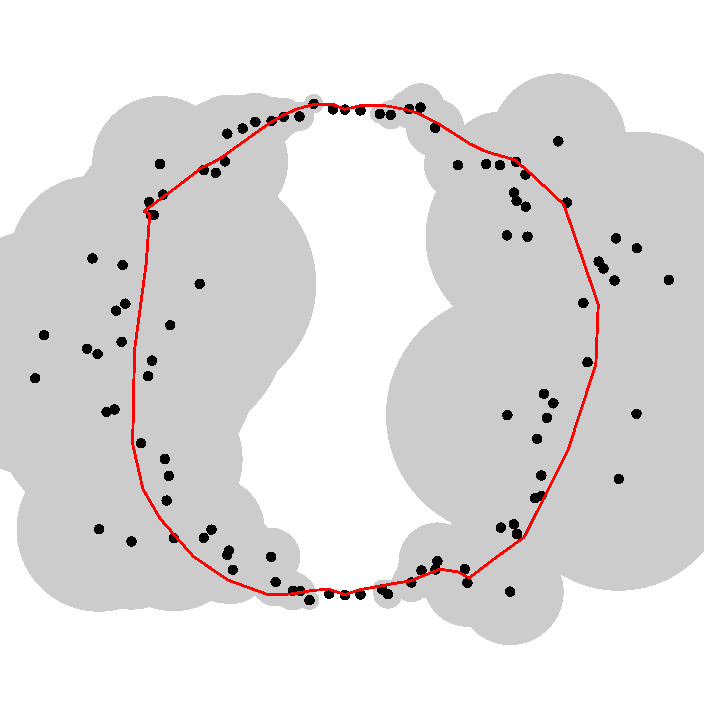}}\hfill
\subfigure[$\delta$=0.75r]{\includegraphics[width=1.4in]{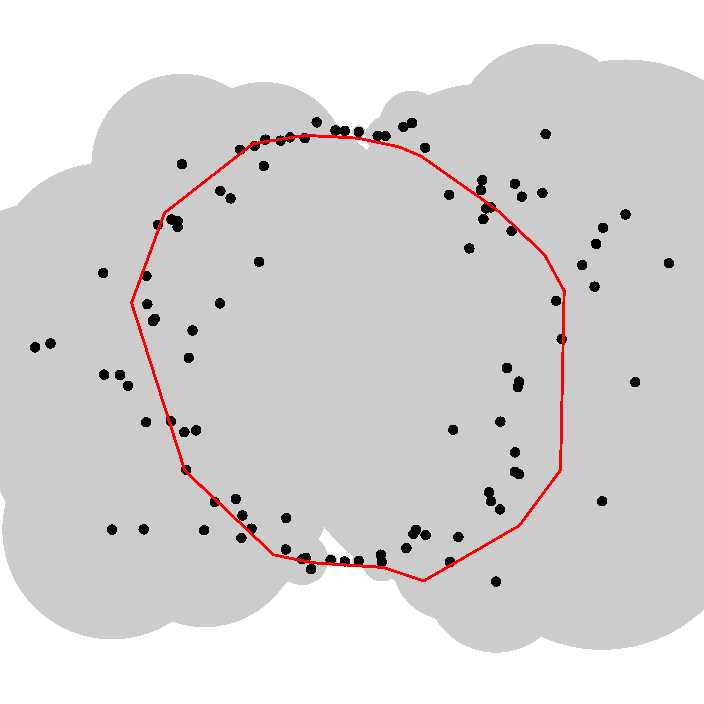}}\hfill
\subfigure[$\delta$=r]{\includegraphics[width=1.6in]{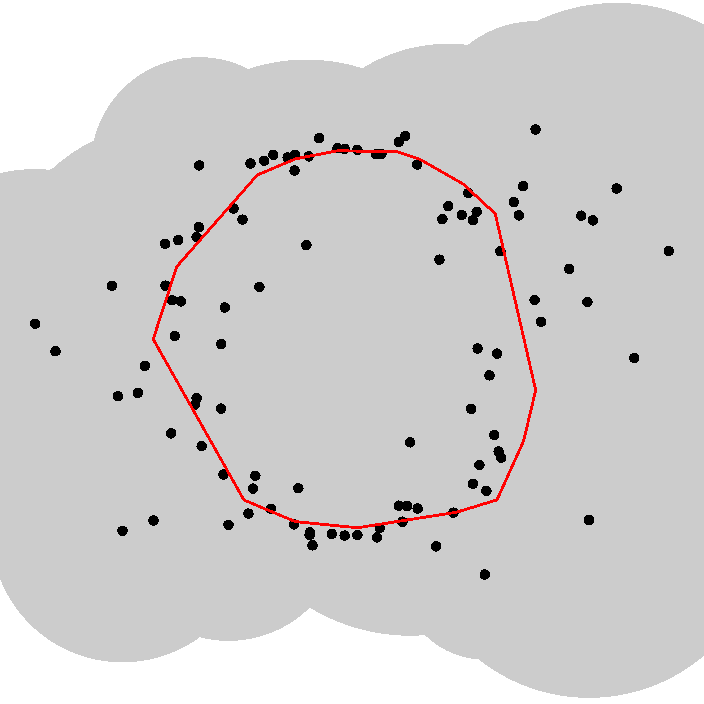}}
\caption{Top: 100 samples on a circle, perturbed with varying (sides: full, top/bottom: zero) noise extent (grey shaded discs) up to $\delta$ of its radius. Row \#2: Connectivity as recovered by {\scshape FitConnect}, Row \#3: Blending of fitted circular arcs as in {\scshape FitConnect} post-processing. Bottom: Our denoising based on the connectivity recovered by {\scshape FitConnect} and taking into account individual noise extents per sample.
}
\label{fig:circles}
\end{figure*}

\begin{table}[h]
  \scriptsize
  \begin{center}
    \begin{tabular}{|l|r|r|r|r|r|r|r|r|r|}
    \hline
    Noise & Input & & & Blend & & & Ours & & \\
    \hline
      $\delta$ & max & mean & RMS & max & mean & RMS & max & mean & RMS \\
    \hline
      0.1 & 0.076 & 0.016 & 0.023 & 0.073 & 0.013 & 0.020 & \textbf{0.023} & \textbf{0.006} & \textbf{0.008} \\
      0.25 & 0.183 & 0.039 & 0.059 & 0.109 & 0.024 & 0.034 & \textbf{0.069} & \textbf{0.020} & \textbf{0.027} \\
      0.5 & 0.367 & 0.079 & 0.117 & \textbf{0.126} & \textbf{0.041} & \textbf{0.053} & 0.140 & 0.042 & 0.055 \\
      0.75 & 0.553 & 0.118 & 0.175 & 0.188 & \textbf{0.053} & \textbf{0.069} & \textbf{0.162} & 0.056 & 0.073 \\
      1 & 0.741 & 0.155 & 0.230 & 0.233 & 0.079 & 0.098 & \textbf{0.145} & \textbf{0.054} & \textbf{0.065} \\
    \hline
    \end{tabular}
  \caption{Comparison of the error of the noisy input samples versus {\scshape FitConnect} blending and our denoising method, as Hausdorff distances from the original circle. The noise varies as shown in Figure~\ref{fig:circles} and all values are in terms of the circle radius.
}
  \label{table:varyingnoise}
  \end{center}
\end{table}

To generate the noise, we use a model that adds uniform random radial noise in the range $[0,\delta]$ with uniform random direction~\cite{mehra2010visibility}.
Figure~\ref{fig:circles} shows how our method is able to recover the circle curve from very large extents of noise (up to its entire radius) and denoise it effectively, compared to simple blending of the fitted circular arcs that {\scshape FitConnect} performs as post-processing.
Table~\ref{table:varyingnoise} shows how well both approaches reduce the input noise, and that our method mostly denoises much better, reducing the input noise (mean or RMSE) typically by a factor of 2-3.

\textbf{Adding samples improves reconstruction}

\begin{figure*}
\centering
\subfigure[]{\includegraphics[width=1.1in]{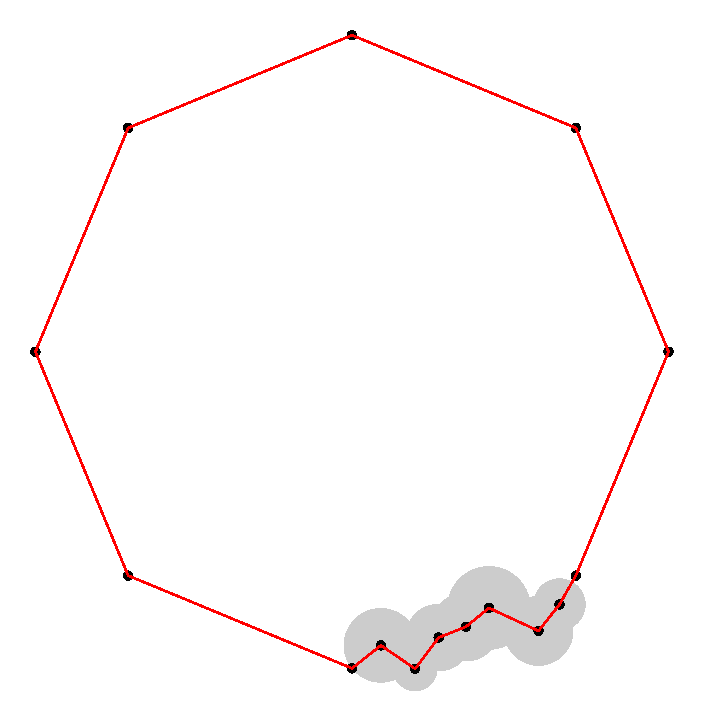}}\hfill
\subfigure[$\epsilon$=0.4, 76 samples]{\includegraphics[width=1.1in]{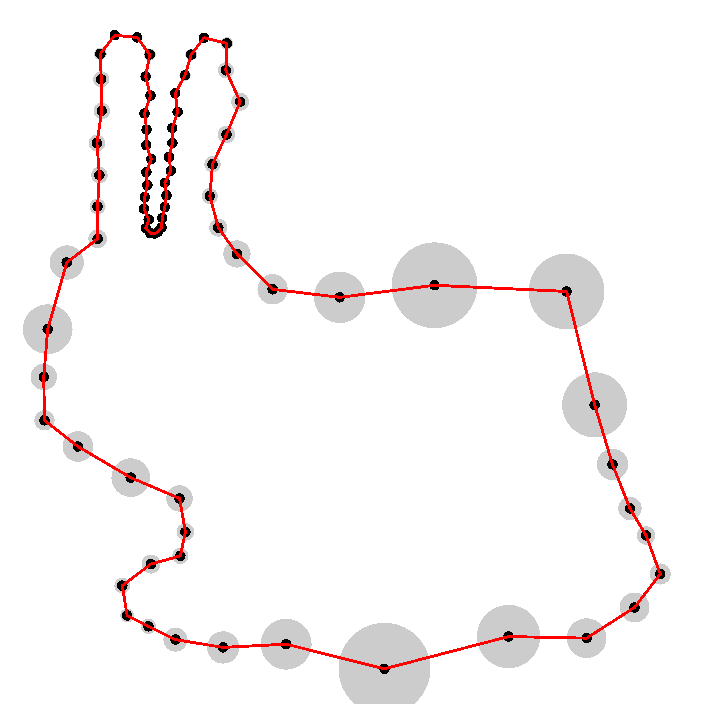}}\hfill
\subfigure[$\epsilon$=0.3, 116 samples]{\includegraphics[width=1.1in]{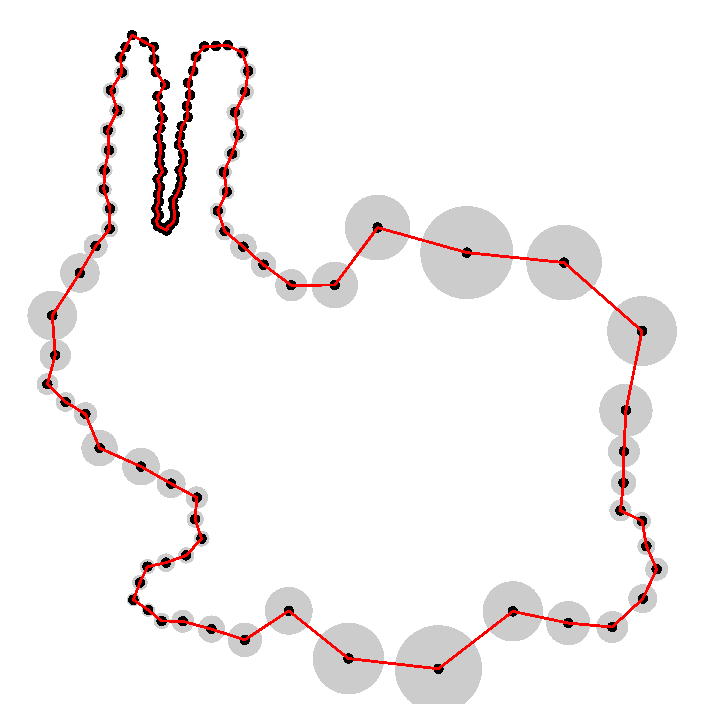}}\hfill
\subfigure[$\epsilon$=0.2, 199 samples]{\includegraphics[width=1.1in]{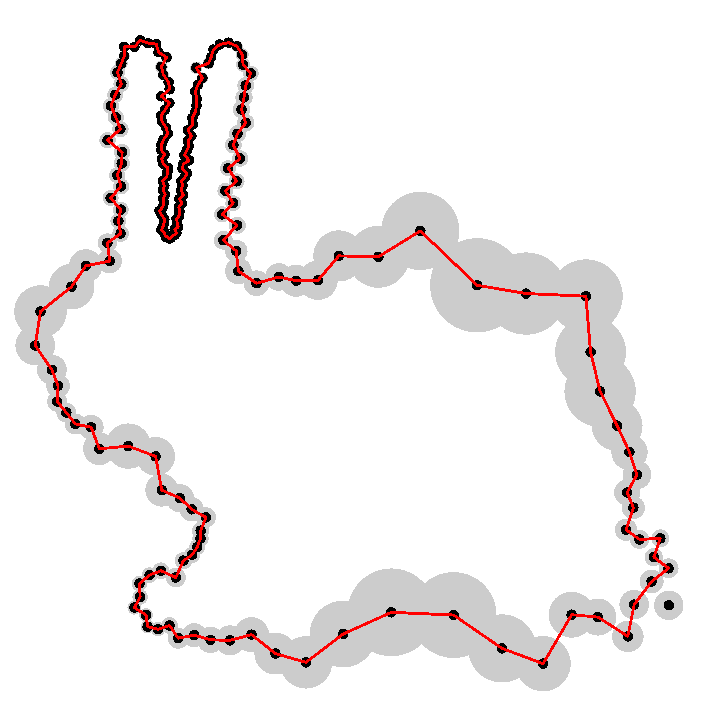}}\hfill
\subfigure[$\epsilon$=0.1, 1464 samples]{\includegraphics[width=1.1in]{fig_density_d1.png}}\hfill
\subfigure[$\epsilon$=0.1, 1464 samples]{\includegraphics[width=1.1in]{fig_bunny_orig.png}}\hfill
\subfigure[]{\includegraphics[width=1.1in]{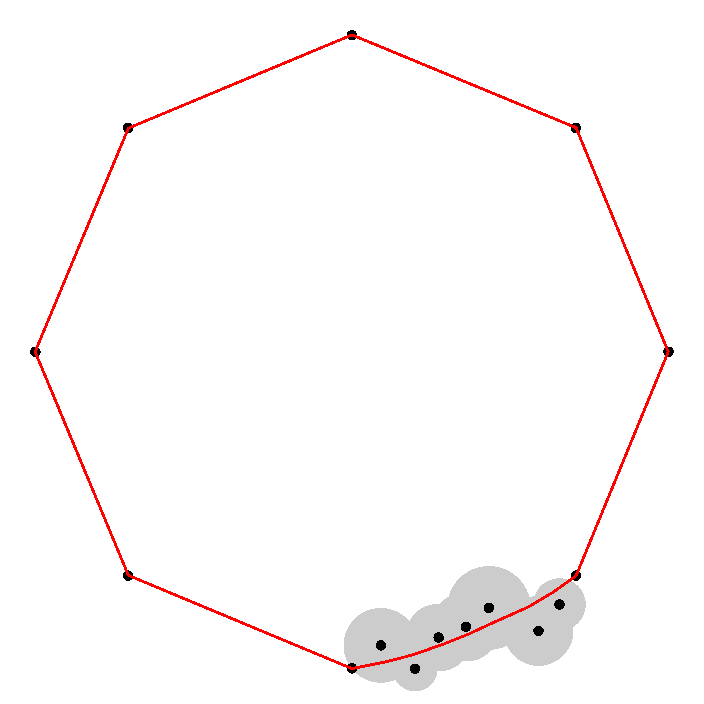}}\hfill
\subfigure[]{\includegraphics[width=1.1in]{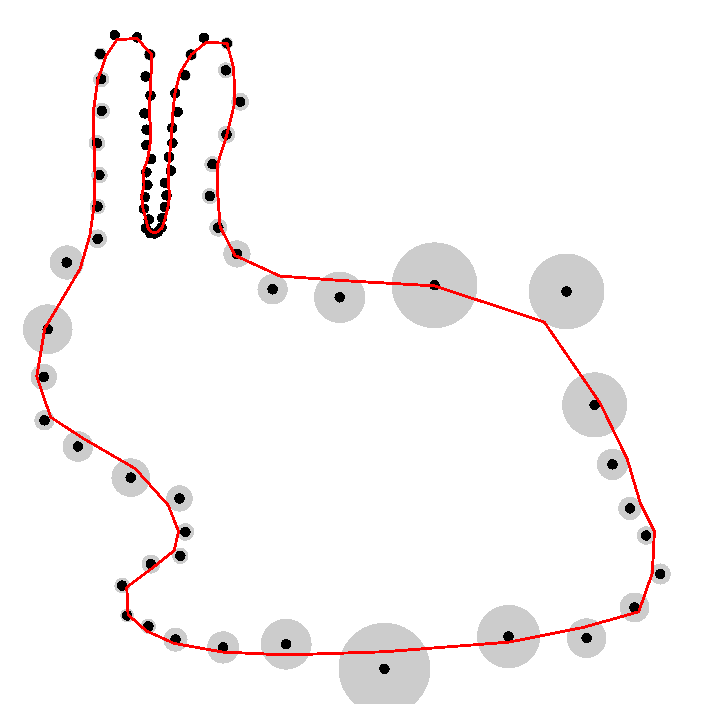}}\hfill
\subfigure[]{\includegraphics[width=1.1in]{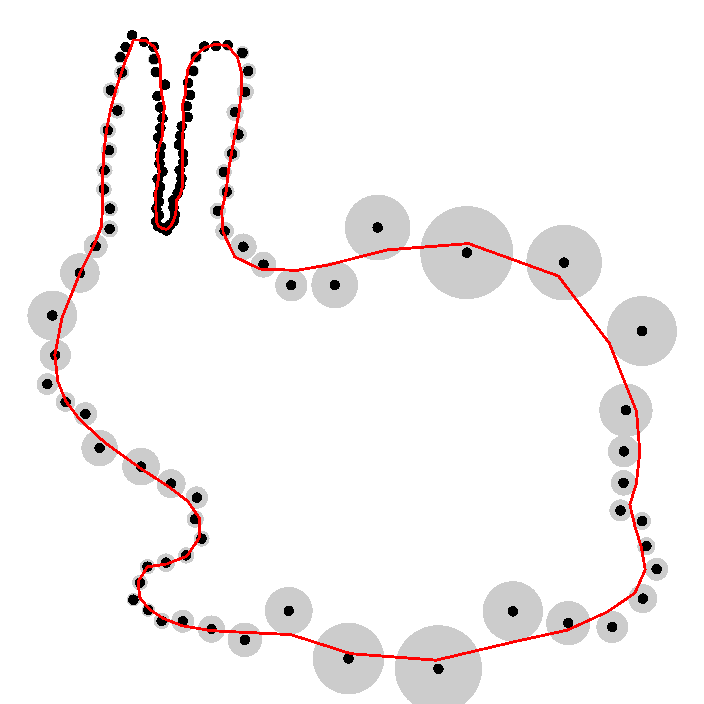}}\hfill
\subfigure[]{\includegraphics[width=1.1in]{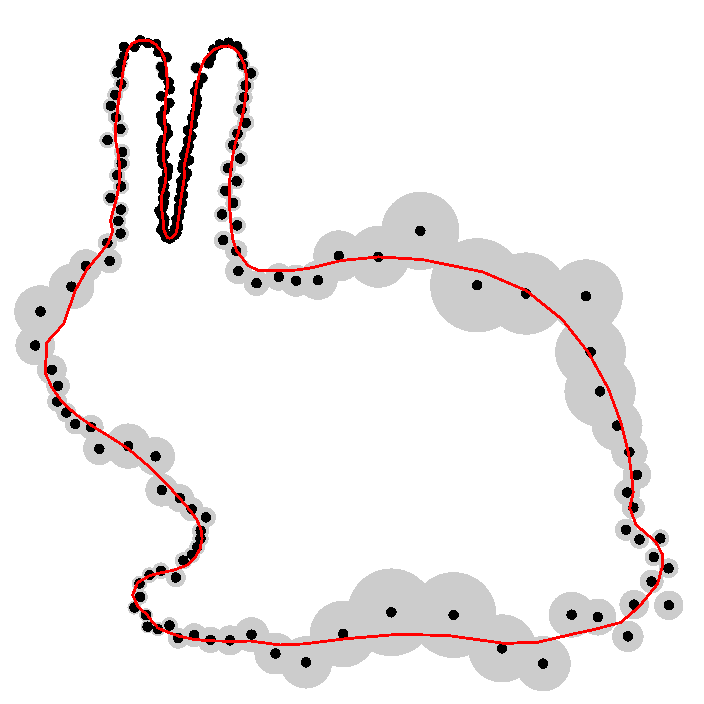}}\hfill
\subfigure[]{\includegraphics[width=1.1in]{fig_density_d2.png}}\hfill
\subfigure[]{\includegraphics[width=1.1in]{fig_bunny_orig.png}}\hfill
\caption{Top: {\scshape FitConnect} connectivity, Bottom: Our denoised output. Left: Noisy subset of circle. Second from left to second from right: {\scshape Bunny} perturbed with noise extent of $\frac{1}{3}$lfs and sampled with increasing density, improving reconstruction quality. Right: Noise-free {\scshape Bunny}.
}
\label{fig:density}
\end{figure*}

Figure~\ref{fig:density} shows that adding more noisy samples to a point set improves the reconstruction quality and that in the limit the reconstruction approaches the original curve.

\textbf{Approximation quality and run time}

\begin{table}[h]
  \scriptsize
  \begin{center}
    \begin{tabular}{|l|r|r|r|r|r|r|r|}
    \hline
      Figure/Noise & \# & old $\overline{SD}$ & new $\overline{SD}$ & old$\sum\angle$ & new$\sum\angle$ & Conn & Den \\
    \hline
      {\scshape Circle} 0.1r & 100 & 0.042 & 0.006 & 2638 & 404 & 0.012 & 0.023 \\
      {\scshape Circle} 0.25r & 100 & 0.091 & 0.233 & 2826 & 595 & 0.062 & 0.011 \\
      {\scshape Circle} 0.5r & 100 & 0.303 & 0.193 & 2396 & 861 & 0.060 & 0.012 \\
      {\scshape Circle} 0.75r & 100 & 0.633 & 0.029 & 689 & 389 & 0.123 & 0.005 \\
      {\scshape Circle} r & 100 & 3.853 & 1.764 & 518 & 326 & 0.142 & 0.003 \\
      {\scshape Bunny} $\epsilon=0.4$ & 76 & 0.000 & 0.093 & 2048 & 1354 & 0.004 & 0.030 \\
      {\scshape Bunny} $\epsilon=0.3$ & 116 & 0.000 & 0.013 & 3461 & 1658 & 0.009 & 0.038 \\
      {\scshape Bunny} $\epsilon=0.2$ & 199 & 0.169 & 0.156 & 8736 & 1845 & 0.022 & 0.060 \\
      {\scshape Bunny} $\epsilon=0.1$ & 460 & 0.821 & 0.787 & 22402 & 5591 & 0.277 & 0.174 \\
      {\scshape Keyboard} & 585 & 0.145 & 0.088 & 13054 & 10162 & 1.243 & 0.150 \\
      {\scshape Monitor} & 915 & 0.037 & 0.199 & 15106 & 12900 & 14.468 & 0.167 \\
      {\scshape Cup} & 263 & 0.838 & 0.886 & 4686 & 3415 & 0.987 & 0.034 \\
      {\scshape Mouse} & 157 & 0.333 & 0.188 & 5323 & 3782 & 0.089 & 0.022 \\
      {\scshape Apple} & 170 & 0.001 & 0.056 & 7951 & 1724 & 0.056 & 0.024 \\
      {\scshape Butterfly} & 164 & 0.124 & 0.091 & 7385 & 1629 & 0.032 & 0.033 \\
      {\scshape Crab} & 284 & 0.331 & 0.279 & 11436 & 4533 & 0.233 & 0.034 \\
      {\scshape Dolphin} & 179 & 0.217 & 0.171 & 8049 & 3015 & 0.080 & 0.033 \\
      {\scshape Fish} & 1000 & 0.666 & 0.322 & 2033 & 1304 & 15.330 & 0.036 \\
      {\scshape Bottle} & 1000 & 0.221 & 0.288 & 2772 & 1475 & 11.334 & 0.029 \\
      {\scshape Bunny} hi noise & 2512 & 0.185 & 0.020 & 11616 & 7075 & 60.645 & 0.186 \\
      {\scshape VarCircle} & 15 & 0.000 & 0.041 & 697 & 362 & 0.002 & 0.003 \\
      {\scshape Square} & 18 & 0.000 & 0.389 & 912 & 392 & 0.002 & 0.003 \\
      {\scshape Sawtooth} & 30 & 0.000 & 0.055 & 1486 & 494 & 0.001 & 0.005 \\
   \hline
    \end{tabular}
  \caption{\# of samples per object, average signed distance ($\overline{SD}$) in \% of point set diagonal as well as total angle sum (deg), each before and after applying our denoising method. Runtime in seconds, for the two passes {\em Conn}ectivity recovery and {\em Den}oising each.}
  \label{table:approximation}
  \end{center}
\end{table}

Table~\ref{table:approximation} shows that our algorithm significantly straightens the curve (the total sum of angles always becomes smaller) and that the curve is well balanced in terms of signed distance to the samples (often much reduced, never becomes large in absolute terms).
The runtime of our unoptimized method is mostly limited by the time taken for the connectivity recovery of {\scshape FitConnect}, which is in principle linear but has quadratic time complexity in the size of noise clusters, while our denoising pass is fast and mostly linear.

\textbf{Comparison with noisy reconstruction algorithms}

\begin{figure*}
\centering
\subfigure[{\scshape Apple}]{\includegraphics[width=1.7in]{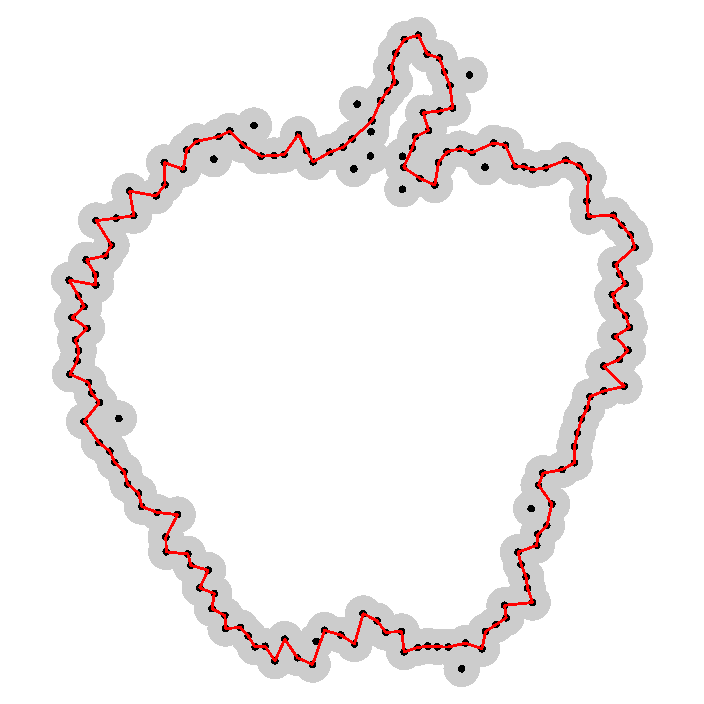}}\hfill
\subfigure[{\scshape Butterfly}]{\includegraphics[width=1.7in]{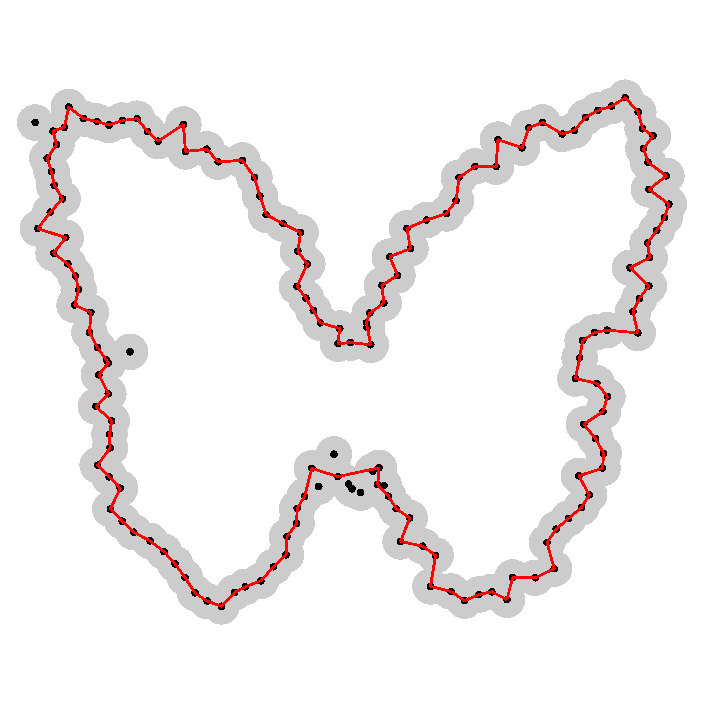}}\hfill
\subfigure[{\scshape Crab}]{\includegraphics[width=1.7in]{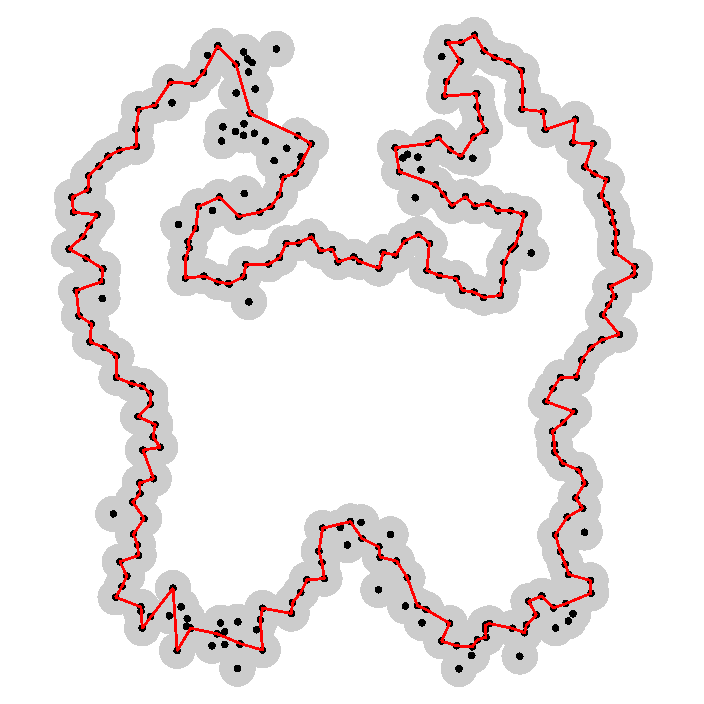}}\hfill
\subfigure[{\scshape Dolphin}]{\includegraphics[width=1.7in]{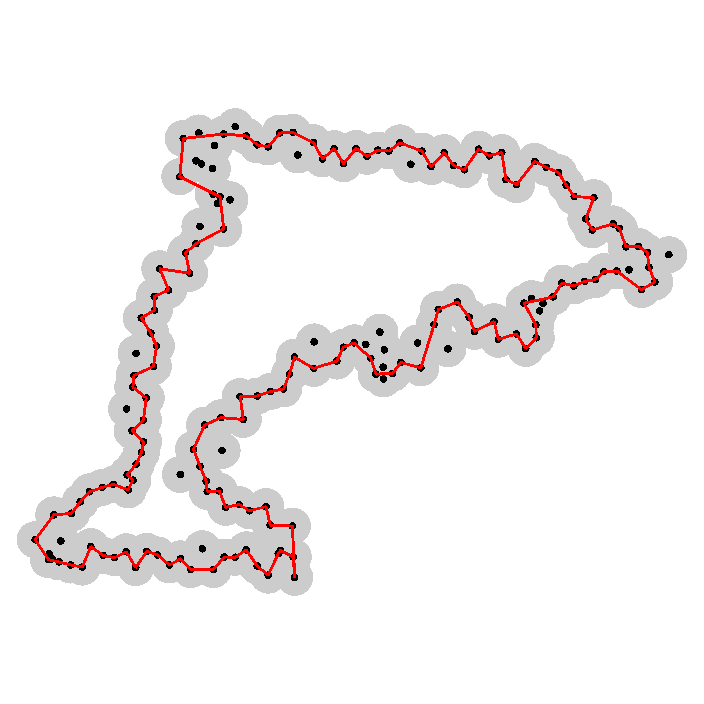}}\hfill
\subfigure[{\scshape Apple}]{\includegraphics[width=1.7in]{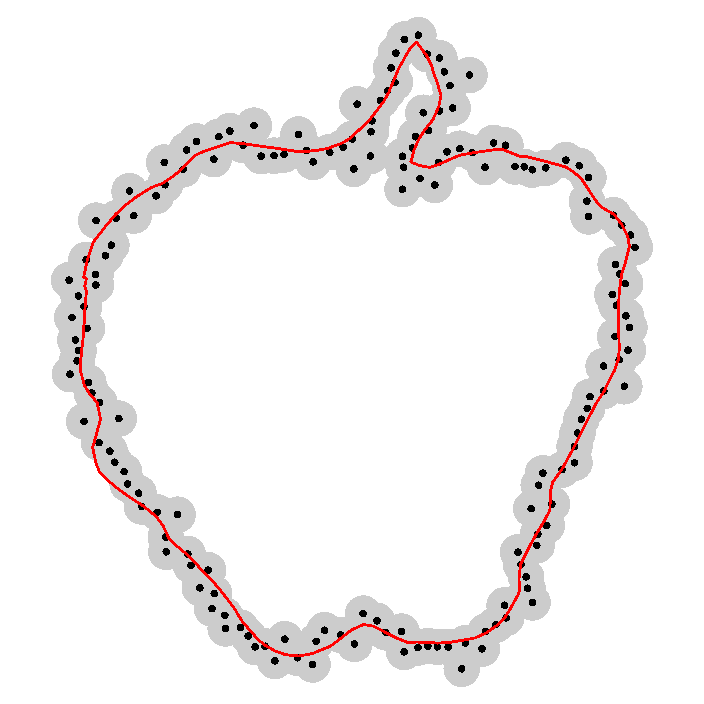}}\hfill
\subfigure[{\scshape Butterfly}]{\includegraphics[width=1.7in]{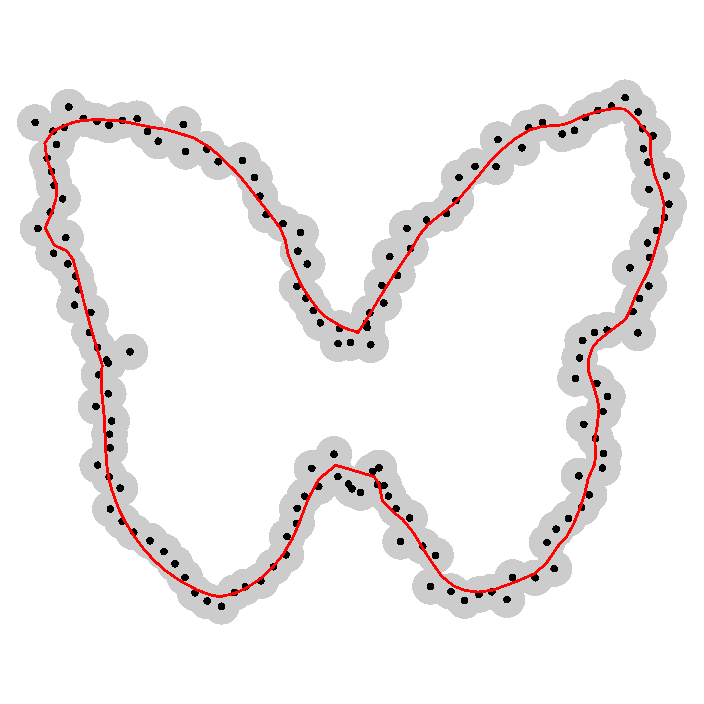}}\hfill
\subfigure[{\scshape Crab}]{\includegraphics[width=1.7in]{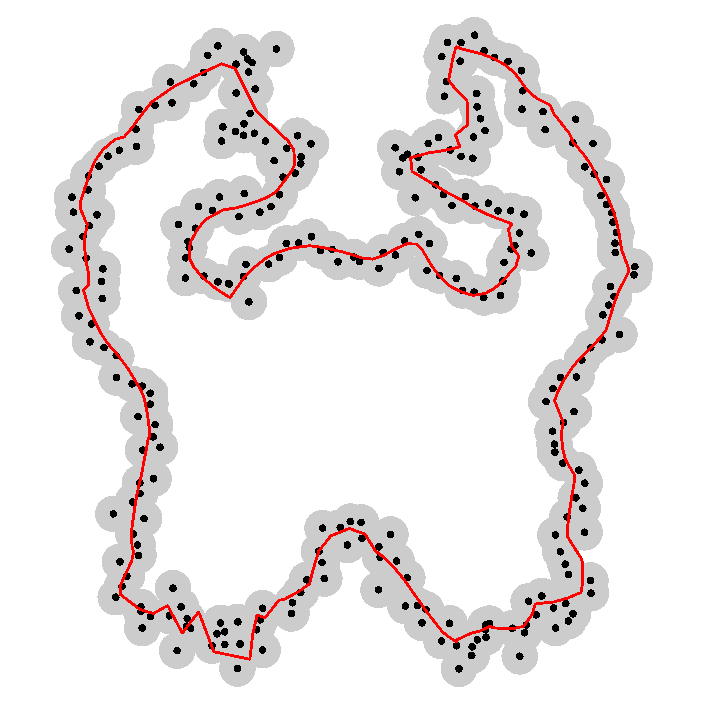}}\hfill
\subfigure[{\scshape Dolphin}]{\includegraphics[width=1.7in]{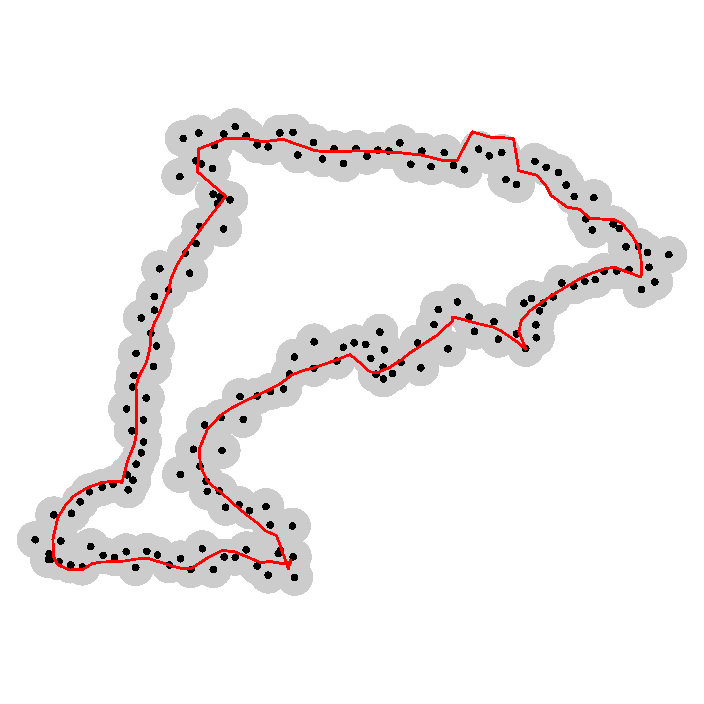}}\hfill
\caption{Reconstruction of point sets which {\em Robust HPR}~\cite{mehra2010visibility} fails to close and denoises only minimally (compare to center column of Fig. 6 in~\cite{mehra2010visibility}): Top: {\scshape FitConnect} connectivity. Bottom: Our manifold and denoised reconstruction for an assumed uniform noise extent.
}
\label{fig:comparison}
\end{figure*}

Figure~\ref{fig:comparison} shows that our method yields better connectivity and denoises much better than {\em Robust HPR} (compare center column in Fig. 6 of~\cite{mehra2010visibility}).

\begin{figure}
\centering
\subfigure[]{\includegraphics[width=1in]{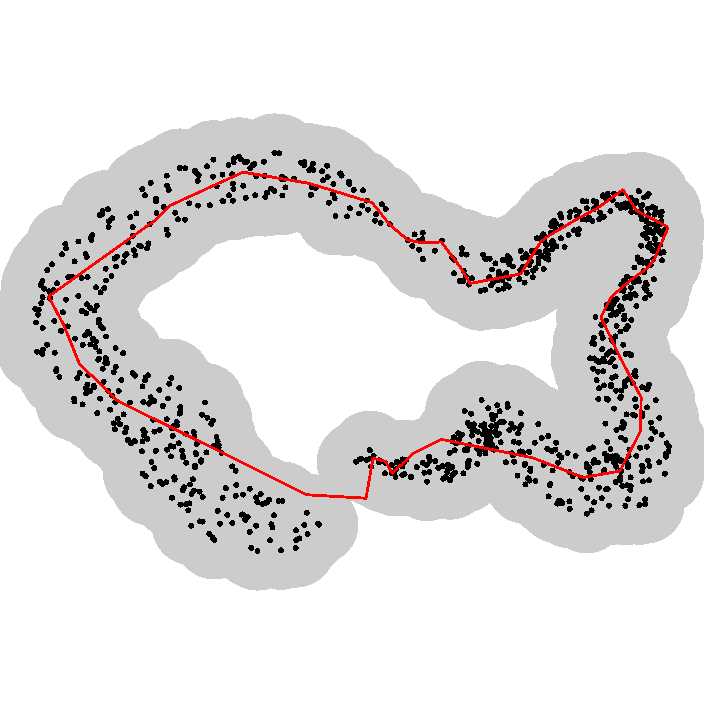}}\hfill
\subfigure[]{\includegraphics[width=1in]{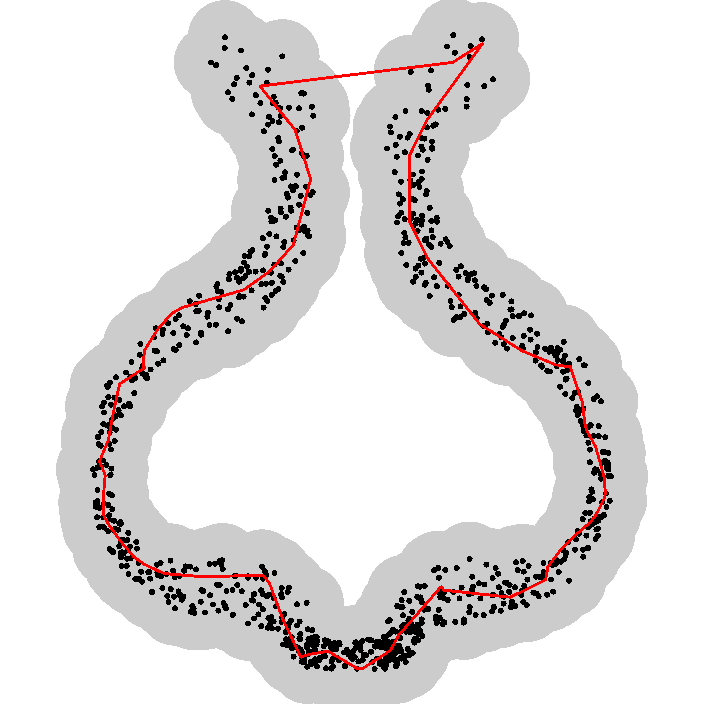}}\hfill
\subfigure[]{\includegraphics[width=1in]{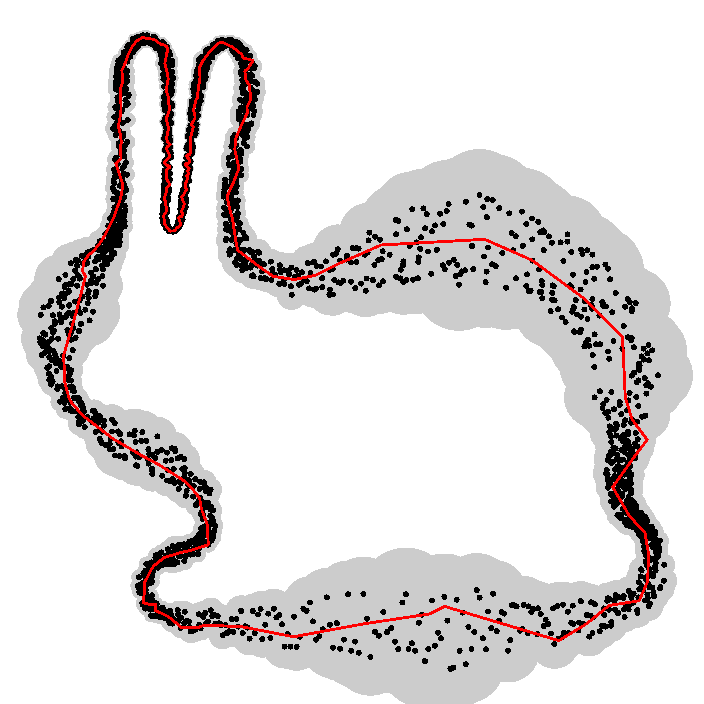}}\hfill
\caption{Reconstruction of highly noisy point sets. Left and center: from a noisy curve construction algorithm (point sets courtesy of Lee~\cite{lee2000curve}), with assumed uniform noise extent. Right: {\scshape Bunny} with approximate noise extent of $\delta=\frac{1}{3}$lfs. 
}
\label{fig:highnoise}
\end{figure}

Figure~\ref{fig:highnoise} shows the results of comparing our denoising method on point sets with uniform very high noise. Note that the compared algorithm only works on open curves whereas {\scshape FitConnect} reconstruction closes the curve (see Fig. 13+14 in~\cite{lee2000curve}). Further, it is iterative as opposed to ours, requires parameter tuning, and while its regression analysis will produce a nice-looking smooth curve, it is likely to over-smooth fine features.

\subsection{Guarantees}

Our method guarantees to preserve all features recovered by {\scshape FitConnect} which protrude over the local noise extent, and a maximum distance to the ground truth curve at vertices (with probability if the distances are given as cut-off radii of a sensor noise model).

\textbf{Feature reconstruction}

\begin{figure}
\includegraphics[width=1.5in]{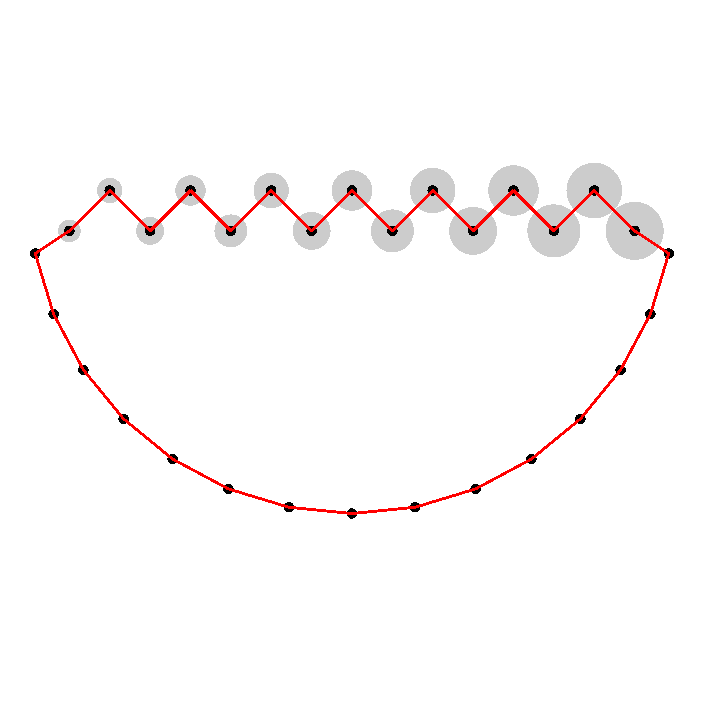}\hfill
\includegraphics[width=1.5in]{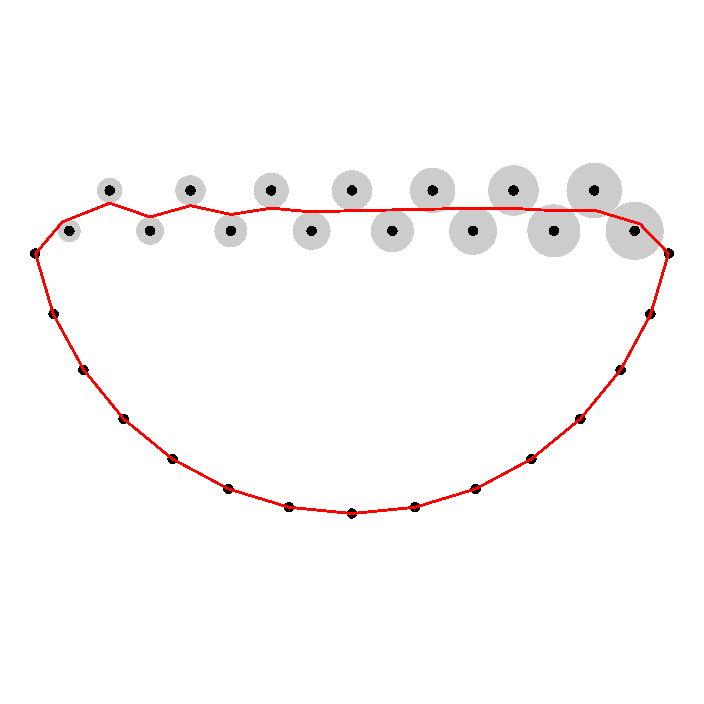}
\caption{{\scshape Sawtooth}: Left half of sawtooth features protrude over the local noise extent and are preserved while right half is merged.
}
\label{fig:sawtooth}
\end{figure}

Figure~\ref{fig:sawtooth} shows a sawtooth configuration of points with increasing amplitude of noise extents.
For the samples left of the center, the noise extent is smaller than the feature size, from the center to the right the noise extent submerges the features.
Consequently, features are preserved for the samples on the left side, while the samples on the right side merge into a single curved segment.

\textbf{Distance to ground truth}
For the synthetic test data above, our denoising method guarantees that the reconstructed curve passes within the specified noise extent of the samples because we limit its movement to these bounds.
For real data, these extents correspond to a stochastic guarantee since the PDF cut-off radius our algorithm considers correlates to a probability value.
We analyze this for real data below.

\subsection{Reconstruction from real data}

\textbf{Silhouettes with estimated noise}

\begin{figure}
\centering
\subfigure[{\scshape Cup}]{\includegraphics[width=1.6in]{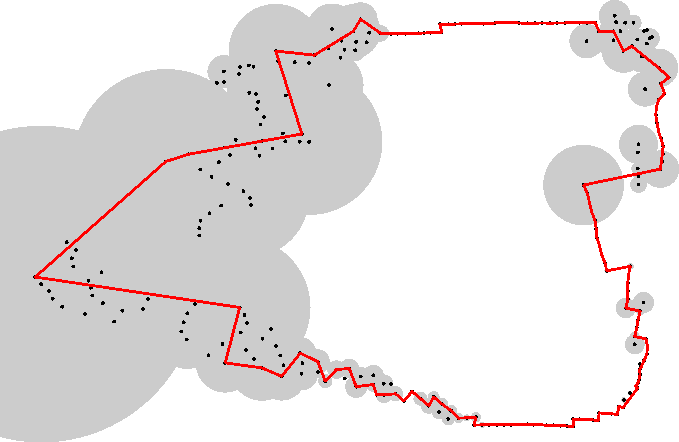}}\hfill
\subfigure[{\scshape Cup}]{\includegraphics[width=1.6in]{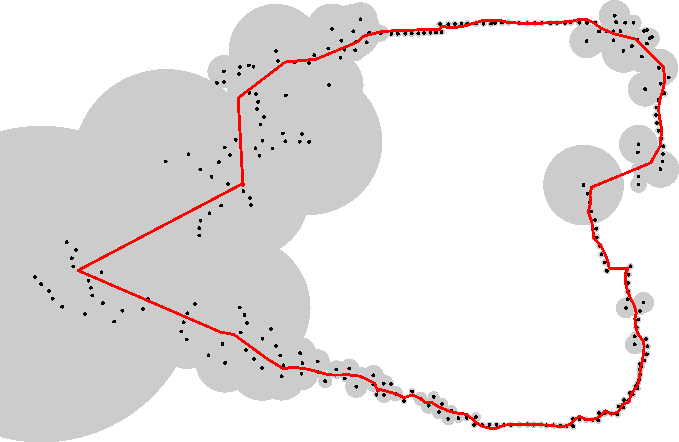}}\hfill
\subfigure[{\scshape Keyboard}]{\includegraphics[width=1.6in]{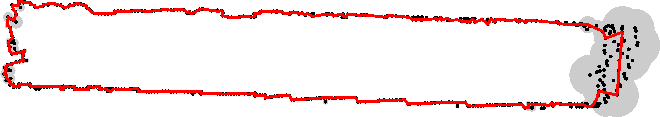}}\hfill
\subfigure[{\scshape Keyboard}]{\includegraphics[width=1.6in]{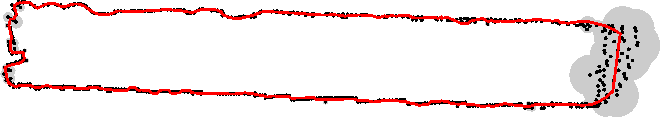}}\hfill
\subfigure[{\scshape Monitor}]{\includegraphics[width=1.6in]{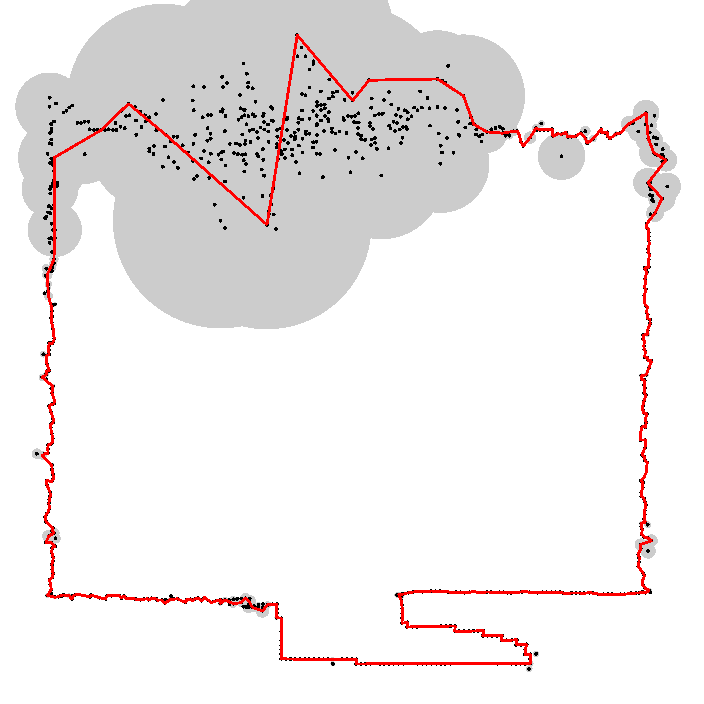}}\hfill
\subfigure[{\scshape Monitor}]{\includegraphics[width=1.6in]{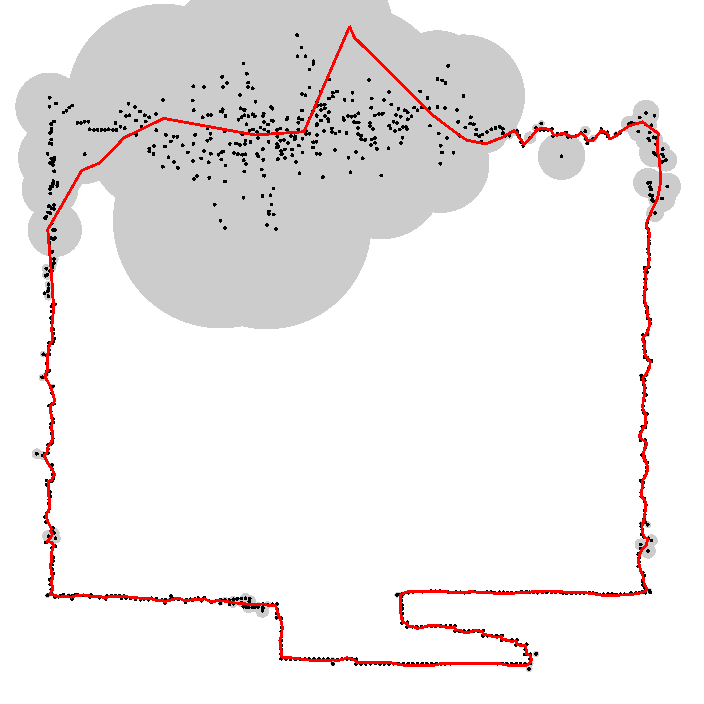}}\hfill
\subfigure[{\scshape Mouse}]{\includegraphics[width=1.6in]{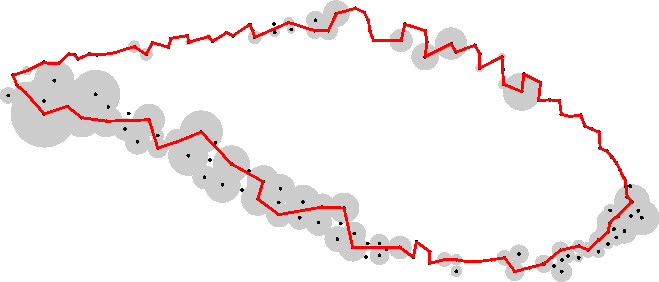}}\hfill
\subfigure[{\scshape Mouse}]{\includegraphics[width=1.6in]{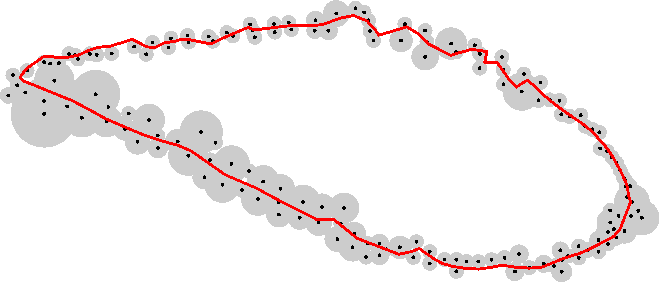}}\hfill
\caption{Segmented silhouettes of sensed 3D objects~\cite{birkas2016mobile} denoised with the individual noise extents per sample detected by {\scshape FitConnect} but a minimum noise extent of constant 1mm since sensor noise properties are not known for these. Left: {\scshape FitConnect} connectivity. Right: Our manifold and denoised reconstruction.
}
\label{fig:silhouettes1}
\end{figure}

\begin{figure}
\centering
\subfigure[{\scshape Drill}]{\includegraphics[width=1.2in]{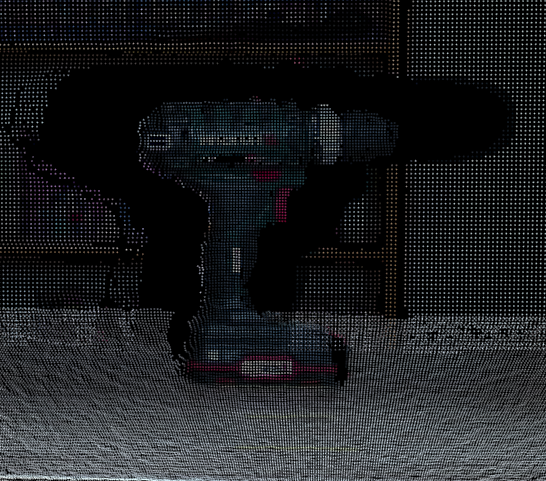}}\hfill
\subfigure[{\scshape Drill}]{\includegraphics[width=0.9in]{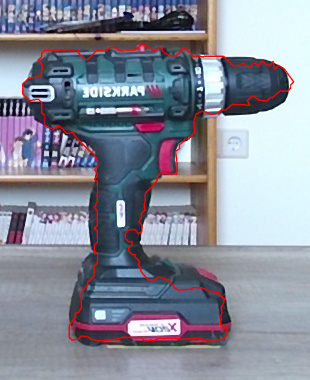}}\hfill
\subfigure[{\scshape Drill}]{\includegraphics[width=0.9in]{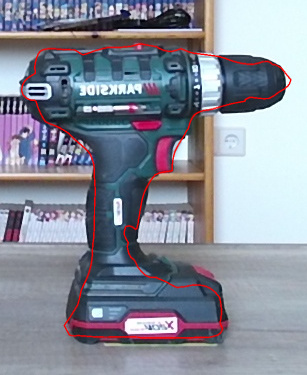}}\hfill
\subfigure[{\scshape Thing}]{\includegraphics[width=1.2in]{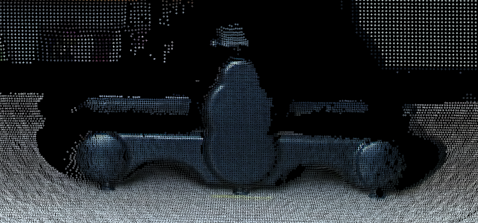}}\hfill
\subfigure[{\scshape Thing}]{\includegraphics[width=1in]{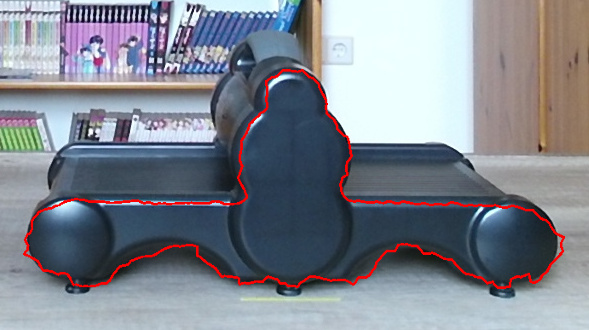}}\hfill
\subfigure[{\scshape Thing}]{\includegraphics[width=1in]{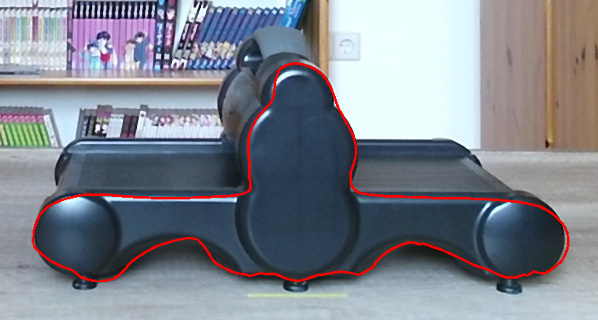}}\hfill
\subfigure[{\scshape Vase}]{\includegraphics[width=1in]{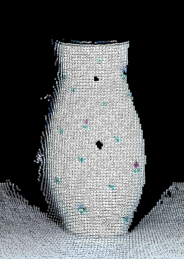}}\hfill
\subfigure[{\scshape Vase}]{\includegraphics[width=0.82in]{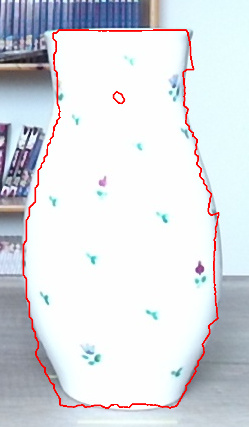}}\hfill
\subfigure[{\scshape Vase}]{\includegraphics[width=0.8in]{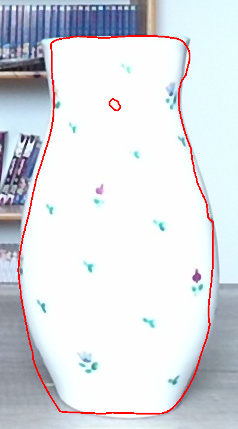}}\hfill
\caption{Segmented silhouettes of sensed 3D objects denoised with the individual noise extents per sample detected by {\scshape FitConnect} but a minimum noise extent of derived from range image properties of the samples. Left: Sensed RGBD point cloud. Center: {\scshape FitConnect} connectivity overlaid on RGB image. Right: Our denoised reconstruction overlaid on RGB image. Note that some deviations are due to imprecise silhouette extraction which is not part of our method.
}
\label{fig:silhouettes2}
\end{figure}

Figure~\ref{fig:silhouettes1} shows segmented silhouettes of sensed 3D objects~\cite{birkas2016mobile}.
Here we use the noise extent estimated by {\scshape FitConnect} for denoising since we do not have information about the actual error from the sensor for these data.
In some (mostly straight) regions with little noise, {\scshape FitConnect} might just interpolate the samples since it foremost tries to preserve features.
That happens because it will detect noise only if the noisy samples are sufficiently densely clustered such that they can be interpolated in a consistent way.
Therefore we set a minimum uniform noise extent of 1mm.

\textbf{Silhouettes with sensor-specified noise}

In Figure~\ref{fig:silhouettes2}, we show segmented silhouettes of sensed 3D objects where the noise extent is computed from the range image properties of the samples' (x,y,z) position.
Note that the extracted silhouettes show some deviations to the objects' real boundaries in the images, due to the used silhouette extraction algorithm.

\subsection{Limitations}

\begin{figure}
\includegraphics[width=1.5in]{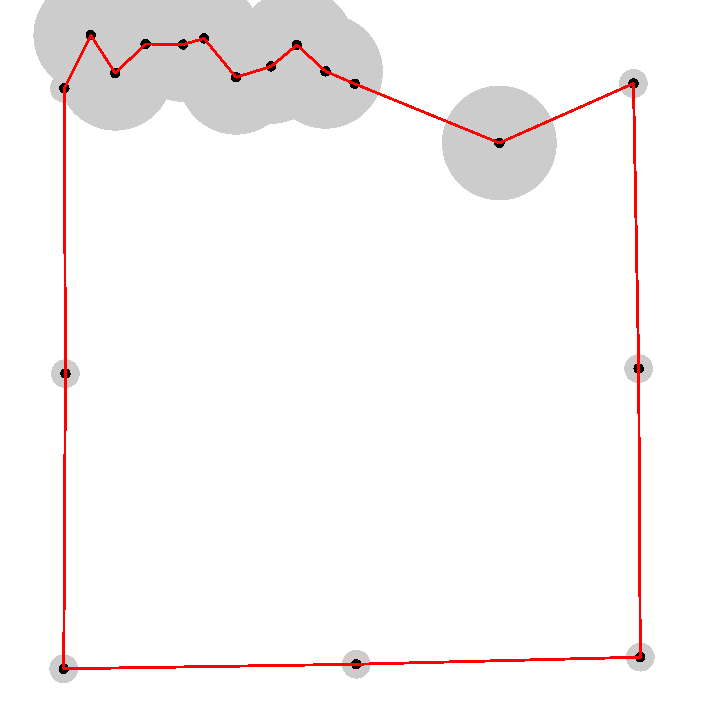}\hfill
\includegraphics[width=1.5in]{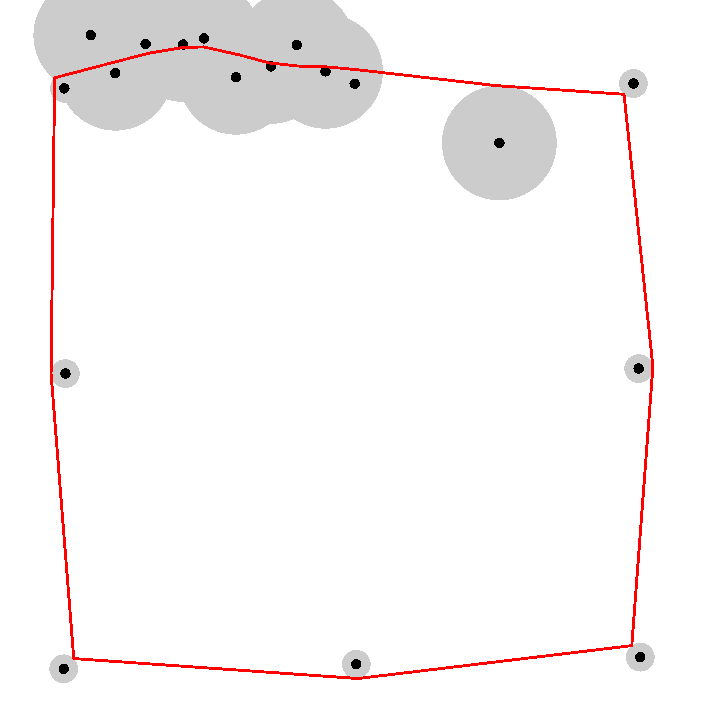}
\caption{{\scshape Square}: straight segments become rounded off.
}
\label{fig:rect}
\end{figure}

Figure~\ref{fig:rect} shows that curves containing straight segments, e.g., silhouettes from man-made objects, are rounded off at their incident corners.
This happens because our objective function minimizes all angles in the least-squares sense and therefore tries to reduce the sharp angles at the corners as much as possible while making in-between straight edges curvy.

\section{Conclusion}

We have shown that our two-pass method successfully enables reconstructing a curve from arbitrarily noisy points within a stochastically guaranteed distance to the original curve while at the same time retaining the features emerging over the local noise extent.
The error between the reconstructed and original curve is guaranteed in terms of the input noise, which can be provided either by sensor-specific properties, or estimates from {\scshape FitConnect}.
Our method is parameter-free since we model the requirements of a most probable curve as minimization, equality and bounds respectively.
We successfully apply a technique that we developed ourselves to solve this constrained optimization problem effectively and efficiently.
One sample application is determining silhouettes of objects in sensed data, however the underlying assumptions extend directly into 3D where reconstruction is a much more interesting and challenging problem.
Our non-optimized denoising algorithm runs fast enough for practical use, it can be verified using the open source available online.

Further extensions aside from reconstruction of surfaces for 3D objects include a sharp corner detector to optimize in-between segments locally, e.g., straight lines of man-made objects, as well as handling open curves.

\section{Acknowledgements}

This work has been funded by the Austrian Science Fund (FWF) project no. P24600-N23.
Data sets {\scshape Keyboard}, {\scshape Monitor}, {\scshape Cup} and {\scshape Mouse} are thanks to Krisztian Birkas.
Data sets {\scshape Drill, Thing, Vase} are thanks to Martin Novak.

\bibliographystyle{eg-alpha-doi}

\bibliography{article}

\end{document}